\documentclass[12pt]{iopart}

\usepackage{graphicx}
\usepackage{amssymb}
\usepackage{epstopdf}
\usepackage{bm} 
\newcommand{\BEQ}{\begin{equation}}
\newcommand{\EEQ}{\end{equation}}
\newcommand{\BEA}{\begin{eqnarray}}
\newcommand{\EEA}{\end{eqnarray}}

\newcommand{\nn}{\nonumber }

\begin{document}

\title[Random, thermodynamic and inverse first order transitions]{Random, thermodynamic and
inverse first order transitions in the  Blume-Capel spin-glass}

\author{Ulisse Ferrari and Luca Leuzzi} \address{Dipartimento di
Fisica, Universit\`a "Sapienza", and IPCF-CNR, UOS Roma, Piazzale Aldo Moro 2,
I-00185, Rome, Italy} \ead{luca.leuzzi@cnr.it}

\begin{abstract}
The spherical mean field approximation of a spin-$1$ model with
$p$-body quenched disordered interaction is investigated. Depending on
temperature and chemical potential the system is found in a paramagnetic
or in a glassy phase and the transition between these phases can be of
different nature.  In given
conditions inverse freezing occurs.  As $p=2$ the glassy phase is replica symmetric and the
transition is always continuous in the phase diagram.  For $p>2$ the exact solution for the glassy
phase is obtained by the one step replica symmetry breaking Ansatz.
Different scenarios arise for both the dynamic and the thermodynamic
transitions. These include (i) the usual random first order transition
(Kauzmann-like) preceded by a dynamic transition, typical of
mean-field glasses, (ii) a thermodynamic first order transition with
phase coexistence and latent heat and (iii) a regime of inversion of static and
dynamic transition lines. In the latter case a thermodynamic stable glassy phase, with
zero configurational entropy,
is dynamically accessible from the paramagnetic phase.   Crossover between different
transition regimes are analyzed by means of Replica Symmetry Breaking
theory and a detailed study of the complexity and of the stability of
the static solution is performed throughout the space of external
thermodynamic parameters.
\end{abstract}

\maketitle

\section{Introduction}
In macromolecular compounds in solution, complex molecules or
polymeric chains can fold into practically inactive conformations,
displaying a negligible interaction with the surrounding system.  The
presence of inactive components can induce the existence of a fluid
phase at a temperature below the temperature range at which the system
is in a solid phase (crystalline, semi-crystalline or amorphous,
depending on the degree of frustration) \cite{P4MP1,aCD,MetCel}. This
corresponds to the occurrence of an {\em inverse transition}, else
said "melting upon cooling", that is a reversible transition between a
completely disordered isotropic fluid phase and a solid phase whose
entropic content is greater than the entropy of the fluid.

This effect can be reproduced and studied in statistical mechanical
 models on a lattice with bosonic spin-$1$ variables, where the holes
 $s=0$ play the role of inactive states. In these models the fluid
 phase is the paramagnet and the solid phase is either a ferromagnet
 (no or weak disorder) or a spin-glass (strong disorder).  A prototype
 model with two-body interactions between bosonic spins is the
 Blume-Emery-Griffiths (BEG) model with ferromagnetic interaction \cite{BEG,
 BC, Schupper04, Schupper05} and its random magnetic implementation
 \cite{Ghatak77,Lage82, Mottishaw85,daCosta94,Arenzon96, Sellitto97,
 daCosta97,Schreiber99, daCosta00, Albino00, Crisanti02, Crisanti04,
 Crisanti05}.

 In presence of quenched disorder the random BEG model is known to
 display both a continuous paramagnet/spin-glass phase transition and
 a first order one.  First order in the thermodynamic sense, i.e.,
 with latent heat and a region of phase coexistence. Furthermore,
 inverse freezing takes place, with a spin-glass at high $T$ and paramagnet
 at low $T$. These properties have been observed in the mean-field
 approximation \cite{Crisanti05}, where the self-consistent solution
 for the spin-glass phase is computed in the full Replica Symmetry
 Breaking (RSB) Parisi Ansatz \cite{Parisi79,Parisi80,MPV}, and on the cubic 3D
 lattice with nearest-neighbor couplings in systems exchanging
 two-body interaction, as well \cite{PLC10,LPC11}. The frustrated BEG model has been studied, as well, by means of numerical  renormalization group techniques with apparently different results: on the 
 Migdal-Kadanoff hierarchical lattice in 3D, it displays no inversion in the transition, nor a low $T$ discontinuity \cite{Oczelik08}, whereas  on the Wheatstone-bridge hierarchical lattice in dimension $\log 12/\log 2 \simeq 3.585$ the inversion seems to be there \cite{Antenucci11}.

Mean-field spin-glass models with more than two-spin interactions,
called $p$-spin models, are known to yield the so-called {\em random}
first order transition, i.e., a phase transition across which no
internal energy discontinuity occurs but the order parameter (the
Edwards-Anderson overlap $q_{\rm EA}$) jumps from zero to a finite
value.  Their glassy phase is described by an Ansatz with one
RSB.  The thermodynamic transition is
preceded (in a cooling procedure) by a dynamic transition due to the
onset of a {\em very large} number of metastable states separated by {\em high}
barriers. The phenomenology of the $p$-spin spin-glass systems is, in
many respects, very similar to the one of structural glasses  \cite{Thirumalai,LeuzziNieuw}. These
models are, therefore, sometimes called {\em mean-field glasses}.

``Very large'' means that the number of states ${\cal N}$ grows exponentially
with the size $N$ of the system: ${\cal N}\sim \exp \Sigma N$, where
the coefficient $\Sigma$ is the configurational entropy, else called
{\em complexity} in the framework of quenched disordered systems. This
is a fundamental property both in mean-field systems \cite{Thirumalai,Bray80,
Crisanti92, Crisanti03, CavagnaPede} and out of the range of validity of mean-field theory, e.g., in
computer glass models \cite{Sciortino05, Binder05}, or, indirectly, by
measuring the excess entropy of glasses in experiments, see, e.g.,
\cite{LeuzziNieuw} and references there in.

``High barriers'' means that the free energy difference between a
local minimum in the free energy functional of the configurational
space (else called {\em free energy landscape}) and a nearby maximum
(or saddle) grows with $N$. That is, it diverges in the thermodynamic
limit. This is a specific artifact of mean-field glasses. The thus
induced dynamic transition corresponds to the transition predicted by
another mean-field theory for the dynamics of supercooled liquids: the
mode coupling theory \cite{Goetze}. The thermodynamic transition
occurring at a lower temperature is, instead, the mean-field
equivalent of the so-called Kauzmann transition in glasses, else known
as {\em ideal glass transition}, predicted by Gibbs and Di Marzio \cite{Gibbs58}.

What happens to mean-field glasses when model features belonging to
systems undergoing inverse freezing are introduced?
The mean-field $p$-spin models are built either using discrete Ising variables
\cite{Gross85,Gardner85}, soft spins \cite{Thirumalai} or spherical
spins \cite{Crisanti92}, the latter two being approximation of
Ising discrete variables that allow for an easier analytic treatment.
 
In this work, we are interested in studying what happens when we mix the
ingredients leading to an inverse transition (the hole state) and to a
random first order transition ($p$-spin interaction), i.e., to provide
a mean-field theory for inverse freezing between fluid and structural glass.
We will see how, in this investigation, 
 non-trivial features will arise, among which the inversion of dynamic
 and static transition and the consequent possibility of accessing low
 energy glassy states without running into dynamic arrest.

\section{Model}
\label{s:model}
The model Hamiltonian that we will consider, derives from 
\begin{equation}
{\cal H}=-\sum_{i_1<\ldots <i_p}J_{i_1\ldots i_p}
s_{i_1}\ldots s_{i_p}
+D\sum_{i}s_i^2
\end{equation}
with $s_i = 1,0,-1$.  The couplings $J$ are Gaussian independent
identically distributed variables with probability distributon:
\begin{equation}
P(J_{i_1\ldots i_p})= \sqrt{\frac{ N^{p-1}}{\pi J^2 p!}} \exp\left\{
-\frac{N^{p-1} J_{i_1\ldots i_p}}{J^2 p!}
\right\} 
\end{equation}
The coefficient $D$, known as crystal field in Blume Capel (BC)
models \cite{BEG,BC}, plays the role of a chemical potential for the empty
sites (holes).

The $p=2$ case was introduced in Ref. \cite{Ghatak77} and it has been
studied subsequently throughout the years
\cite{Lage82, Mottishaw85,daCosta94,Arenzon96, Sellitto97,
 daCosta97,Schreiber99, daCosta00, Albino00, Crisanti02, Crisanti04,
 Crisanti05}. A preliminary study of the $p$-body case was
presented in Ref. \cite{Mottishaw86}.  

The spin-$1$ case can be translated into an Ising spin problem on
lattice gas. Indeed, if we rewrite $s_i = \sigma_i
n_i$ and crystal field $D = D'+T\log 2$, with $\sigma=\pm 1$ and $n=0,1$
we have 
\begin{equation}
{\cal H} = - \sum_{i_1<\ldots <i_p} J_{i_1\ldots i_p}
\sigma_{i_1}\ldots \sigma_{i_p}n_{i_1}\ldots n_{i_p}
+(D'-T\log 2)\sum_{i}n_i
\label{f:Ham}
\end{equation}
We stress that the shift in chemical potential is necessary to keep
the relative degeneracy of zero and non zero values of the spins in
the partition sum \cite{Griffiths67}.  A different shift  in $D$ (or none)
would, actually, define another model. Generally speaking a chemical
potential transformation of the kind $D=D'-T\log 2r$ will provide the
Ising spin on lattice gas model corresponding to a model with spin$-1$
variables taking values $s=1$ (or $s=-1$) $r$ times more frequently than value
$0$.  E.g., $r=1/2$ corresponds to $s=1,0,0,-1$; $r=1$ to the original
case $s=1,0,-1$, cf. , Eq. (\ref{f:Ham}); $r=4$ to $s=1,1,0,-1,-1$ and so
on \cite{Schupper04}. In the following we will address the
variants with generic $r$, as well.

In the Ising $p$-spin model, as $p>2$, besides a random first order
transition (RFOT) between a paramagnet and a spin-glass phase in the 
1RSB Ansatz, also a lower temperature
phase transition is expected to occur between the latter and a full
RSB spin-glass phase.  Indeed, this is what is known to occur in the
$D\to -\infty$ limit \cite{Gardner85}.  Since we are exclusively
interested in the transition between fluid and (mean-field) glass,
thus 1RSB stable, we can simplify the computation by approximating the
discrete spins with continuous real variables satisfying a
global {\em spherical} constraint.

As suggested in Ref. \cite{Caiazzo02}, we introduce, to this aim, the
variable $\tau = \sigma ( 2 n -1) = \pm 1$, such that $n=(\sigma \tau
+1)/2$, and eventually obtain the model Hamiltonian:
\begin{equation}
\hspace*{-2cm}
{\cal H}= - \frac{1}{2^p}\sum_{i_1<\ldots< i_p} J_{i_1\ldots i_p}
(\sigma_{i_1}+\tau_{i_1})\ldots (\sigma_{i_p}+\tau_{i_p})
+(D-T\log 2)\sum_{i=1}^N\frac{\sigma_i\tau_i+1}{2}
\label{f:leadHam}
\end{equation}
with
\BEQ
\sum_{i=1}^N \sigma_{i}^2=\sum_{i=1}^N \tau_i^2 = N.
\EEQ
Applying RSB theory \cite{Parisi79,Parisi80,MPV}
we are going to investigate  thermodynamics and complexity 
of the disordered model represented in
Eq. (\ref{f:leadHam}).

\section{Replicated free energy and order parameters}
\label{sect:rep_f_en}

The replicated free energy, averaged over the distribution of
disordered couplings, reads:
\begin{eqnarray}
f_n & = & - \frac{1}{ n N \beta} \log  \int {\cal D}{\bm Q}{\cal D}{\bm T}{\cal D}{\bm R} \quad \exp \left\{ - n N \beta G(\beta,D;{\bm Q},{\bm T},{\bm R})\right\}
\label{f:FreeEnMatr}
 \\  \nonumber 
-n \beta G &=&  -(\beta D-\log2)  \sum_a\frac{R_{aa} +1}{2} + \frac{\beta^2}{4^{(p+1)}} \sum_{ab} \left( Q_{ab} + T_{ab} + 2R_{ab} \right)^p  +\nonumber \\
&& + \frac{1}{2}\ln \det \left( \frac{1}{2}({\bm {QT}}+{\bm {TQ}})-{\bm R}^2 \right)+ \frac{n}{2}\log \pi + n ,  \nonumber
\\
\nonumber
&& {\cal D} {\bm Q}= \prod_{a<b}^{1,n}dQ_{ab}
\qquad{\cal D}{\bm T} = \prod_{a<b}^{1,n}T_{ab}
\qquad{\cal D}{\bm R} = \prod_{a\leq b}^{1,n}R_{ab}
\end{eqnarray}
where we have inserted the three overlap matrices
\begin{eqnarray}
N Q_{ab} = \sum_i \sigma_i^a \sigma_i^b,  \qquad
N T_{ab} = \sum_i \tau_i^a \tau_i^b, \qquad
N R_{ab}= \sum_i \sigma_i^a \tau_i^b 
\label{f:Q}
\end{eqnarray}
and, consequently, integrated over the spin variables.
The saddle point, self-consistency, equations are:
\begin{eqnarray}
&&Q_{ab} = T_{ab} = R_{ab} , \quad a\neq b \label{f:sceq1} \\
&&-\frac{1}{2} \left[ \left( {\bm Q} {\bm T} - {\bm R}^2  \right)^{-1} R \right]_{ab} =  \frac{\beta D - \log2}{4} \delta_{ab} -\frac{p \beta^2}{2^{p+3}} \left( Q_{ab} + R_{ab} \right)^{p-1}  
\label{f:sceq2}
\end{eqnarray}

From the free energy, cf. Eq. (\ref{f:FreeEnMatr}), in the zero replica limit
\begin{equation}
f=\lim_{n\to 0}G(\beta,D,{\bm Q}^{\rm sp}, \bm{T}^{\rm sp}, \bm{R}^{\rm sp})
\end{equation}
 one derives all
thermodynamic quantities, such as the density of filled-in sites
\begin{equation}
d= \frac{d f}{d D} =  \lim_{n \rightarrow 0} \frac{1}{n}\sum_a\frac{R_{aa} +1}{2};
\end{equation}
the internal energy
\begin{equation}
u = \frac{d \beta f}{d \beta} = - D d  - \lim_{n \rightarrow 0} \frac{2\beta}{4^{(p+1)}}\frac{1}{n} \sum_{ab} \left( Q_{ab} + T_{ab} + 2R_{ab} \right)^p
\end{equation}
and the entropy
\begin{eqnarray}
 s&=& d \log2 + \frac{1}{2}\log \pi + 1 \nonumber \\
&& + \lim_{n \rightarrow 0} \frac{1}{2n}
 \left[ \frac{2 \beta^2}{4^{(p+1)}} \sum_{ab} \left( Q_{ab} + T_{ab} + 2R_{ab} \right)^p+ 
  \ln \det \left(\bm{QT}-\bm{R}^2 \right) \right]. 
\end{eqnarray}

In order to compute the above expressions the precise {\em shape} of
matrices $Q$, $T$ and $R$ has to be identified. There is no {\em a priori} method to
deduce the correct form and one has to resort to an {\em Ansatz}.

The simplest one is the Replica Symmetric (RS) Ansatz, where the discrete
symmetry group $S_n$, of permutation between replicas, holds for all
two-indeces quantities. This means that the elements of each overlap
matrix $M$ only take one value $y$ outside the diagonal and $M$
can be parametrized as:
\begin{equation}
M_{ab} = (M_d-M_0) \delta_{ab} + M_0,
\end{equation}
where $\delta$ is the Kronecker delta. Replica symmetry always holds
 for one index parameters, because physical properties of a single
 replica must be identical for each replica.  Indeed, all single
 index quantities, like the diagonal part of the $R$ overlap matrix,
 are index independent.

RS is not always self-consistent, though. 
Studying the fluctuations in the
space of replica matrices around a RS solution for
Eqs. (\ref{f:sceq1})-(\ref{f:sceq2}) one finds that, for $p>2$, the
glassy solution is not stable and one has to break the symmetry
between replicas in order to find a self-consistent solution. 
 To perform such stability analysis one needs the Hessian matrix.
For our model it is computed in detail in \ref{app:Hessian}.  We will
later report about the stability analysis of both the $p=2$ case and
the $p>2$.  Breaking the Replica Symmetry means to allow for
different values in the off-diagonal elements of the overlap
matrices. The symmetry can be broken step by step allowing a further
overlap value at each step and organizing replicas in clusters with a
hierarchical structure \cite{Parisi79,Parisi80}. For our purpose one step turns out to be
sufficient, leading to a \textit{one step Replica Symmetry Breaking}
(1RSB) solution, in which a generic matrix takes the form:
\begin{equation}
M_{ab} = (M_d-M_1) \delta_{ab} + (M_1-M_0) \epsilon_{ab} + M_0,
\end{equation}
where $\epsilon_{ab}=1$ if $a$ and $b$ belong to the same cluster of size $m$ and
$\epsilon_{ab} =0$ otherwise.

\section{Thermodynamics for $p=2$. Inverse freezing.}
\label{s:p=2}

 For $p=2$ the free energy evaluated in the RS Ansatz reads:

\begin{eqnarray} 
\hspace*{-2cm}- 2\beta \lim_{n \rightarrow 0}  G^{RS} = - 2 (\beta D- \log2) d + \frac{\beta^2}{2}(d^2-q^2) + \frac{q}{\eta} 
+ \log(\theta \eta) + \log (4 \pi) +2  
\label{f:FreeEnRS}
\end{eqnarray}
where $\theta = 1-d$ and $\eta = d-q$ are eigenvalues of the matrix $QT-R^2=Q^2-R^2$. Self-consistency equations read:
\begin{eqnarray}
\beta^2 q &=& \frac{q}{\eta^2}  
\label{f:sceq1_p2_RS}
\\
  \beta^2 \eta&=& \frac{\eta - \theta}{\theta \eta} + 2 ( \beta D - \log 2)
\label{f:sceq2_p2_RS}
\end{eqnarray}

The RS solution turns out to be marginally stable with respect to fluctuations in 
the space of replica parameters, as shown in  \ref{app:RS_stab}.
In particular the lowest relevant eigenvalues of the Hessian,  the so-called replicon, 
cf. Eq. (\ref{f:Replicon_p2}),
is $\Lambda_1^{(1)}=0$.

\begin{figure}[!t]
\center
\includegraphics[height=.26\textheight]{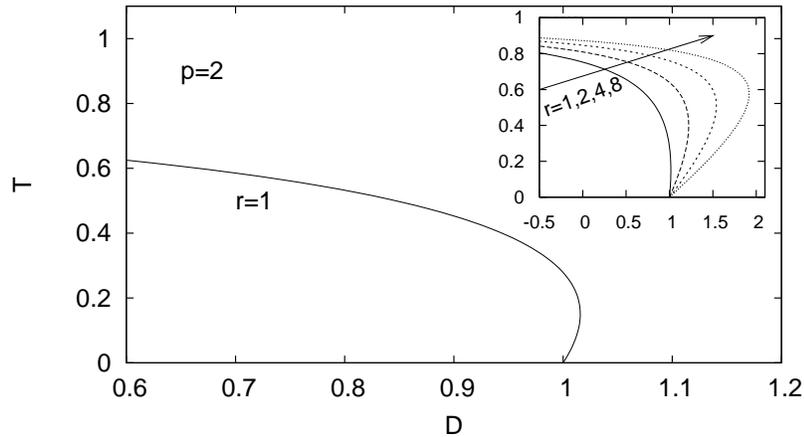}
\caption{$T,D$ phase diagram of the $p=2$ model case. Inset: the same
diagram is shown for models with different values or the parameter
$r=1,2,4,8$, ratio of the number of non-zero to zero values for the
spin in the discrete counterpart of our spherical model. As $r$
increases the inverse transition region is enhanced. }
\label{fig:PhDi_TD}
\end{figure}

\begin{figure}[t!]
\center
\includegraphics[height=.26\textheight]{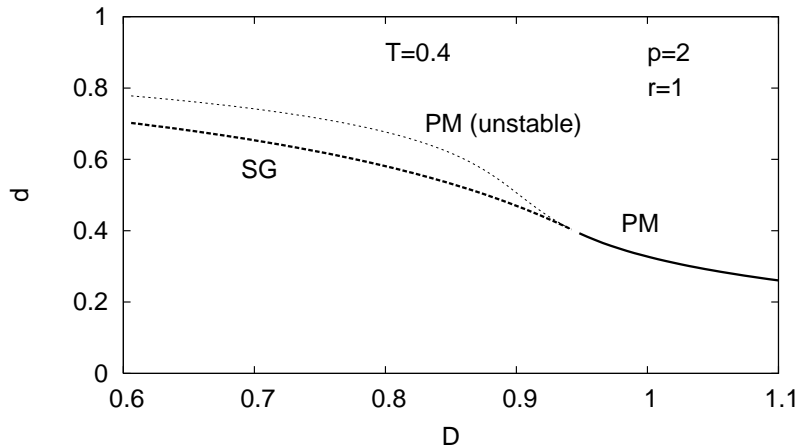}
\caption{Density behavior at $T=0.4$ across the SG/PM transition for the $p=2$ model.  The paramagnetic density is shown (as a thin curve) also in its unstable phase.}
\label{fig:rho_D}
\end{figure}

The self-consistency equation (\ref{f:sceq1_p2_RS}) admits $q=0$ as a
solution. This leads to a paramagnetic phase, characterized by a
density $d_{PM}$, determined imposing equation
(\ref{f:sceq2_p2_RS})\footnote{Actually, there is a region in the
$D-T$ diagram with three solutions for $d_{PM}$, but only one turns
out to be stable. See also Eq. (\ref{f:sceq2_p_PM}) with $p=2$.} 
and with a free-energy:
\begin{eqnarray}
 - \beta f^{RS}_{PM}(\beta,D) &=& - (\beta D- \log2) d_{PM} +
\frac{\beta^2}{4}d_{PM}^2 + \nonumber\\ &&+ \frac{1}{2}\log
[(1-d_{PM})d_{PM}]+ \frac{1}{2}\log (4 \pi) +1
\label{f:FreeEnPM} 
\end{eqnarray}
For low $T$ and $D$ the paramagnetic phase turns out to be unstable,
in particular Eq. (\ref{f:Replicon_p2}) becomes negative and a new
solution to Eqs. (\ref{f:sceq1_p2_RS})-(\ref{f:sceq2_p2_RS}) occurs
with $q>0$:
\begin{eqnarray}
 d(T,D) &=& 1 - \frac{T}{2(1-D + T\log2)} \\
 q(T,D) &=& 1 -T~\frac{\frac{3}{2}-D + T\log2}{1-D + T\log2} .
\end{eqnarray}
This SG solution is stable in the RS Ansatz in the whole region of the
phase diagram where it exists, delimited by the transition line:
\begin{equation}
 D_{c}(T)= 1+ T\log2 - \frac{T}{2(1-T)}.
\end{equation}
In Fig. \ref{fig:PhDi_TD} we plot the phase diagram for the model.
At the transition the overlap parameter $q$ grows continuously from
$0$,
 the paramagnetic and the spin-glass solutions coincide and the
density crosses continuously the transition, as shown in Fig. \ref{fig:rho_D}.
 This defines a second
order transition, without latent heat or discontinuities in first
order derivatives. As shown in the figure, a reentrance of the
transition line points out the presence of inverse freezing.

As we mentioned above, in Sec. \ref{s:model}, one can easily switch
model using Schupper-Shnerb variables \cite{Schupper04,Schupper05}
controlling the degeneracy ratio $r$ between filled and empty sites
in the original model with discrete variables. The effect of
increasing $r$ is to increase the breadth of the interval in chemical
potential for which inverse freezing takes place. In the inset of
Fig. \ref{fig:PhDi_TD} we show the phase diagrams for some choices
of $r$.

\begin{figure}[t!]
 \begin{center}
 \includegraphics[height = .26\textheight]{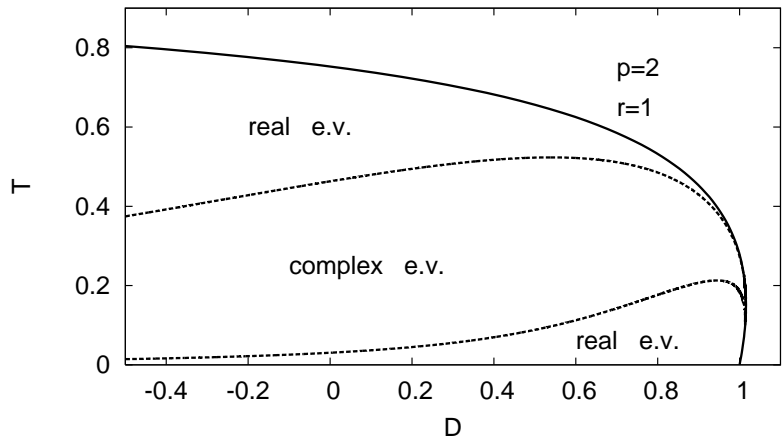}
 \caption{ $T$, $D$ phase diagram: in the inner region between at least  one couple of complex eigenvalues is present in the Hessian in the space of replica fluctuations, as computed on the RS solution. In the other regions all eigenvalues are real.}
 \label{fig:imaginary_ev}
\end{center}
\end{figure}

Finally, we stress that, besides $\Lambda_1^{(1)}$, cf. Eq. 
(\ref{f:Replicon_p2}), there are six more distinct eigenvalues of the
stability Hessian, four of which are strictly real and positive. The
other two, due to the $n\to 0$ limit, can be complex conjugated and develop an imaginary
part. Their real part is always positive. We refer to
 \ref{app:RS_stab}, for a detailed discussion. Here we only stress
that, since the RS solution is exact for $p=2$, the onset of
imaginary eigenvalues is not a signature of lack of consistency of the replica Ansatz but an
artifact due to the limit $n\to 0$ in the replica calculation.  In
Figure \ref{fig:imaginary_ev} we plot the regions of the phase diagram in which imaginary
stability eigenvalues occur.

\subsection{Random matrix approach}
To conclude the analysis of the $p=2$ model we mention that, as shown
 in Refs.  \cite{Caiazzo02,Caiazzo04a,Caiazzo04b}, the thermodynamics
 can be computed also with the method applied by
 Kosterlitz, Thouless and Jones \cite{Kosterlitz76} to the spherical
 Sherrington-Kirkpatrick model, that is the $D\to- \infty$ limit of
 the present model.  The method consists in describing the model in
 terms of the variables $\hat \sigma_i^\lambda$ diagonalizing the
 interaction matrix: 
 \begin{equation}
 \sum_j J_{ij} \hat \sigma_j^\lambda = J_\lambda
 \hat \sigma_i^\lambda
 \end{equation}
 The partition function turns out to be:
\begin{equation}
 Z=\int dz \quad \pi^{\frac{N}{2}}e^{ N \left\{ \frac{\beta }{2}(2z + D - T \log2) - \log\frac{\beta}{2} - \frac{1}{2N} \sum_\lambda \log [\Lambda_1 (z) \Lambda_2(z)] \right\} }.
\label{WIGps}
\end{equation}
where:
\begin{eqnarray}
 \Lambda_1 (z)&=& \frac{\beta}{2} \left(z- \frac{J_\lambda}{2} - \frac{D - T \log2}{2} \right)\\
 \Lambda_2 (z)&=& \frac{\beta}{2} \left( z+\frac{D - T \log2}{2}  \right).
\end{eqnarray}
In the thermodynamic limit, the sum over the eigenvalues can be
evaluated as an integration having as measure Wigner's semicircle law:
\begin{equation}
  \frac{1}{N}\sum_\lambda f(\lambda) \longrightarrow \int_{-2}^{2} dJ \rho(J) f(J) = \int_{-2}^{2} dJ    \frac{1}{2 \pi} \sqrt{4-J}  f(J)
\end{equation}
Our replica calculation agrees with the result of Caiazzo {\em et al.}  \cite{Caiazzo02,Caiazzo04a,Caiazzo04b} .

\section{Static results for $p>2$}
The situation is far richer, and interesting, when more than two-body
interactions among spin-variables are considered.  As $p>2$, indeed,
not only is the RS solution inconsistent in the SG phase (cf. stability analysis in
 \ref{app:RS_stab}) but the exact one step RSB solution yields
different kinds of transitions, some of which atypical in the
framework of mean-field glasses.
 For $p>2$ the 1RSB free energy reads:
\begin{eqnarray}
 - \beta G^{1RSB} &=& -(\beta D - \log2)d + \frac{\beta^2}{4}\left( d^p +(m-1)q_1^p \right)+ \\
&&+\frac{1}{2}\left(2\log2+\log\theta+ \frac{m-1}{m}\log\eta_0 + \frac{1}{m}\log \eta_1  \right), \nonumber
\label{f:FreeEnRSB}
\end{eqnarray}
where 
we have used the expression of the eigenvalues of the $(Q^2-R^2)$  matrix:
\begin{eqnarray}
 \theta &=& 1-d \\
 \eta_0 &=& d - q_1\\
 \eta_1 &=& d + (m-1)q_1 -m q_0
\end{eqnarray}
We set $q_0=0$, because of the absence of an external magnetic field.
Self-consistency equations read:
\begin{eqnarray}
\phi(q_1)&=&\frac{q_1}{\eta_0 \eta_1}  \label{f:sceq1_p_RSB} \\
\phi(d) - \phi(q_1)&=& \frac{\eta_0 - \theta}{\theta \eta_0} + 2 (\beta D -\log2)  \label{f:sceq2_p_RSB} \\
z(y) & =& \frac{2}{p}. 
\label{f:m_static}
\end{eqnarray}
where 
\begin{eqnarray}
\phi(q)&=& \frac{p \beta^2}{2}q^{p-1}\\
y &\equiv& \frac{\eta_0}{\eta_1}
\label{f:ydef}
\end{eqnarray} 
and  
\begin{equation}
z(y) \equiv - 2 y \frac{1- y + \log y}{(1-y)^2}
\end{equation}
is the  Crisanti-Sommers function \cite{Crisanti92}.
The analysis of the stability of the 1RSB solution is reported in Sec. \ref{sec:stab}
and in  \ref{app:1RSB_stab}.

The complexity functional, defined as the Legendre transform of the
free energy functional, evaluated on the saddle point solution yielded
by equations (\ref{f:sceq1_p_RSB}) and (\ref{f:sceq2_p_RSB}),
logarithmically counts the number of metastable states \cite{Bray80}. It reads:
\begin{eqnarray}
\Sigma_{LT}(m,D,\beta) &=&  \beta m^2 \frac{\partial }{\partial m}G(m,D,\beta,q_1^{\rm sp}(m,D,\beta),d^{\rm sp}(m,D,\beta)) \\ \nonumber
&=& 
- \frac{m^2 \beta^2 \left(q_1^{\rm sp}\right)^p}{4} - \frac{m \beta q_1^{\rm sp}}{2\eta_1^{\rm sp}}
 -\frac{\beta}{2} \log \frac{\eta_0^{\rm sp}}{\eta_1^{\rm sp}} 
\label{f:complexity}
\end{eqnarray}
To determine the dynamic transition line $T_d(D)$ one has to impose a
maximal complexity (versus $m$) condition:
\begin{equation}
 \frac{\partial \Sigma_{LT}}{\partial m} = 0 \qquad \leftrightarrow \qquad \eta_1 = (p-1) \eta_0
\label{MAX_COMPL}
\end{equation}
in place Eq. (\ref{f:m_static}).

From the analysis of the paramagnetic solution ($q_0=q_1=0$), solving the self-consistency equation
\begin{eqnarray}
\frac{p}{2}d^{p}(1-d) &=& T^2 (2d-1) + 2 (T D - T^2\log 2)(1-d)d  \label{f:sceq2_p_PM}
\end{eqnarray}
one finds a region in the $T,D$ phase diagram where three solutions for the PM density $d$ occur.
One of them is always unstable, whereas the other two coexist between the spinodal lines. The latter can be
expressed, in a parametric form in $d$, by
\begin{eqnarray}
D =\sqrt{\frac{p}{p-1}}~\frac{d^{-1+p/2} \left(d+p-3 d p+2 d^2 p+2 (1-d)^2 d (p-1) \ln (2r ) \right)}{2 (1-d) 
\sqrt{4d^2-4d+1}}
\\
T =\sqrt{p(p-1)} ~ \frac{d^{p/2} (1-d)}{\sqrt{4d^2-4d+1}}
\end{eqnarray}
The first order transition line is, eventually, obtained by comparing the free energies of the two solutions $d_{\rm PM^+}$ and $d_{\rm PM^-}$.

\begin{figure}
\center
\includegraphics[height=.26\textheight]{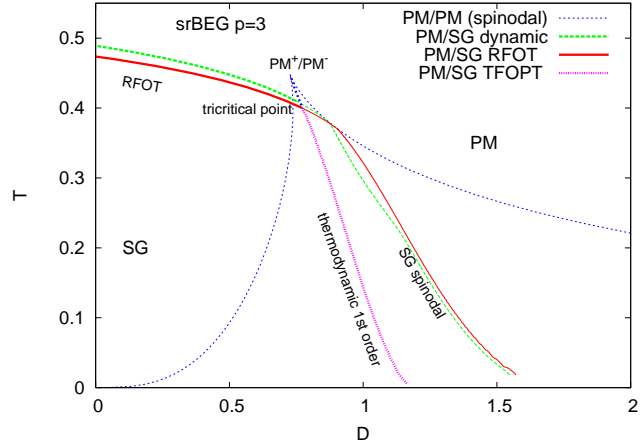}
\caption{$T$,$D$ phase diagram of the spherical $p=3$-spin BC model with filled-in to empty sites ratio $r=1$. }
\label{fig:PhDi_p_TD}
\end{figure}

\begin{figure}
\center
\includegraphics[height=.26\textheight]{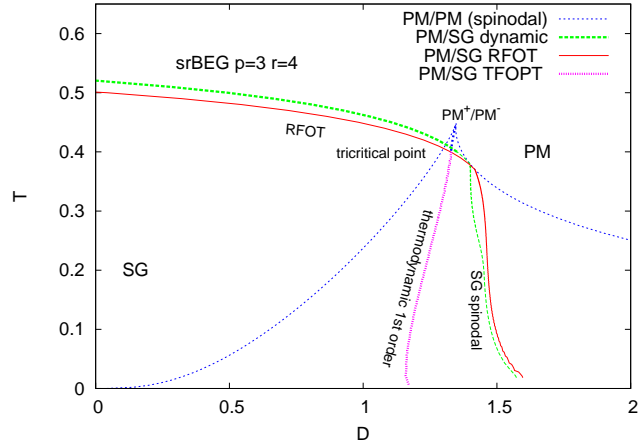}
\caption{$T$,$D$ phase diagram of the spherical $p=3$-spin BC model with  filled-in to empty sites ratio $r=4$.
A reentrance occurs in the first order thermodynamic transition line.}
\label{fig:PhDi_p_TD_r4}
\end{figure}

In Figs. \ref{fig:PhDi_p_TD} and  \ref{fig:PhDi_p_TD_r4} we show the phase diagram for $p=3$, 
both for $r=1$ and $r=4$. For $r=1$ no reentrance of the thermodynamic transition line appears, as shown in Fig. \ref{fig:PhDi_p_TD},
but, increasing $r$, inverse freezing is clearly reproduced, cf. Fig.   \ref{fig:PhDi_p_TD_r4}.
The first order PM$^-$/PM$^+$ line is the thick-dotted line and spinodal lines are plotted as thin-dotted lines.
Qualitatively distinct transition regimes are identified.  We will
dedicate a subsection to each one of them, starting from the most
conventional one.  As a reference we follow the phase diagram in
Fig. \ref{fig:PhDi_p_TD} moving from small (and negative) to large
(positive) values of the chemical potential of the holes.


\subsection{Random First Order Transition (RFOT)}
\begin{figure}[!t]
\center
\includegraphics[height=.26\textheight]{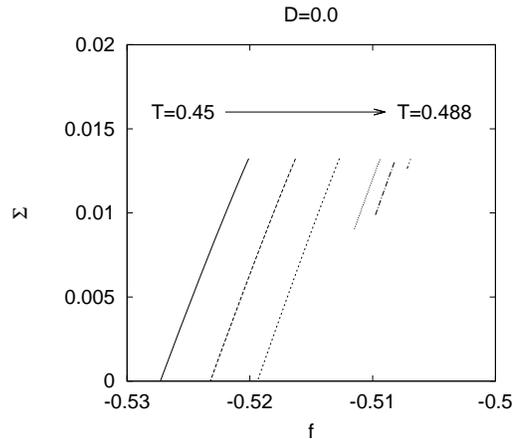}
\caption{$\Sigma(f;T,D=0)$ in a region of temperature around static random first order ($T_s(D=0)= 0.4739$) and dynamic ($T_d(D=0)=0.4892$) transition. }
\label{fig:SigmaMin_T_D}
\end{figure}

For $D$ small enough our model displays the same behaviour as the
spherical $p$-spin model \cite{Crisanti92} and is
an example of the well know RFOT. At high temperature the system is
in a ergodic paramagnetic phase. As it is cooled down, crossing a
temperature $T_d$ it undergoes  a dynamical arrest, despite the
static order parameter $q_1$ is still null. Only at a lower
temperature $T_s<T_d$ a thermodynamic phase transition
occurs. Critical static and dynamic lines are determined solving the
equation systems (\ref{f:sceq1_p_RSB}), (\ref{f:sceq2_p_RSB}),
(\ref{f:m_static}) or (\ref{f:sceq1_p_RSB}), (\ref{f:sceq2_p_RSB}),
(\ref{MAX_COMPL}), respectively.  For $p=3$, $r=1$, e.g., in the $D,T$ phase
diagram they read
 \begin{eqnarray}
D_s(T) &=& \frac{1.8794 T^{1/3}-2.49052 T-0.989769 T^{5/3}}{1 -1.42793 T^{2/3}}
\label{f:TK_D}
\\
D_d(T)&=& \frac{21.6337 T^{1/3} - 27.6822 T - 11.5344 T^{5/3}}{12. - 16.6407 T^{2/3}}
\label{f:Td_D}
\end{eqnarray}
These curves are the $m=1$-lines for statics  and dynamics, respectively
represented in Figs. \ref{fig:PhDi_p_TD}, \ref{fig:PhDi_p_TD_r4} by the thick full (dark grey) and dashed (light grey) lines.
As one can observe in Fig. \ref{fig:SigmaMin_T_D}, crossing the
dynamical temperature from high $T$ values, a collection of high
complexity SG states arise at high free energy values. Each one of
these states
has a free-energy strictly greater than the free energy of the
equilibrium state, namely the paramagnetic one. Indeed, from a
strictly thermodynamic point of view, the stable state is the
paramagnetic one, leaving each SG state as metastable. Decreasing further
the temperature the system gains new states with lower free-energy and
complexity, until, at $T=T_s$, zero-complexity SG states appear, which
turn out to yield the stable SG phase.

\begin{figure}[t!]
\center
\includegraphics[height=.26\textheight]{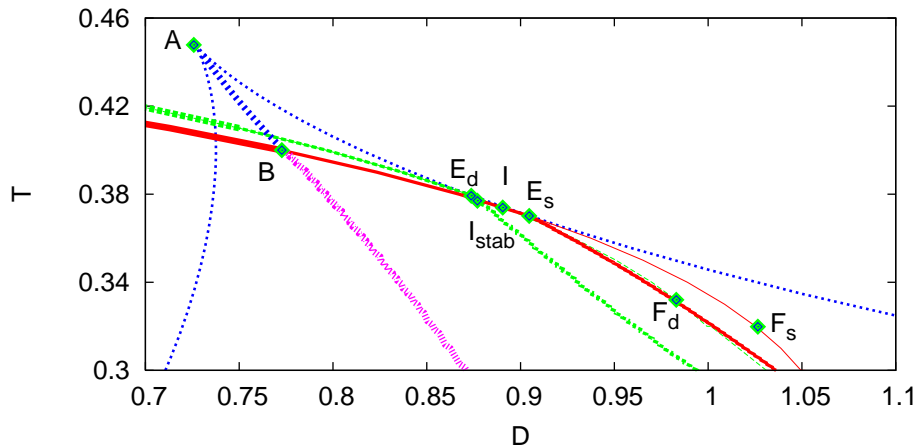}
\caption{Detail of the $D,T$ phase diagram with $r=1$ with all significant critical points. $A$ is the  critical point of the
PM$^+$/PM$^-$ transition; $B$ is the merging point of the first order PM$^+$/PM$^-$ with the first order
SG/PM$^-$ line; $E_{d,s}$ is the limit of stability of the SG (dynamic/static) solution along the $m_{d,s}=1$ line; $I$ is the point of dynamic-static inversion of the $m=1$ lines (where the 1RSB solution is unstable) and $I_{\rm stab}$ is the point of dynamic-static inversion in which the SG solution is stable; $F_{d,s}$ are the points of dynamic-dynamic and static-static $m$-lines crossing, in the region of replica instability.}
\label{fig:punti}
\end{figure}

\subsection{Thermodynamic First Order Phase Transition}
At $(D_A,T_A) \simeq (0.73,0.44)$ a first order transition line starts
developing in the PM phase, separating two distinct PM phases as $T$
is lowered. The two phases are caracterized by their
density, which has a jump crossing the transition line.
We name PM$^{\pm}$ the paramagnetic phase at higher/lower density.
At $(D_B,T_B)\simeq (0.77, 0.40)$
the first order PM$^+$/PM$^-$ transition line merges with a first
order SG/PM$^-$ transition line $D_c(T)$, cf. Fig. \ref{fig:punti}, and the thermodynamic
transition to the SG is not "random" anymore but, rather, a standard
first order, ruled by the Clausius-Clapeyron equation, cf., e.g., Ref. \cite{Crisanti05}.

\begin{figure}[b!]
\center
\includegraphics[width=.47\textwidth]{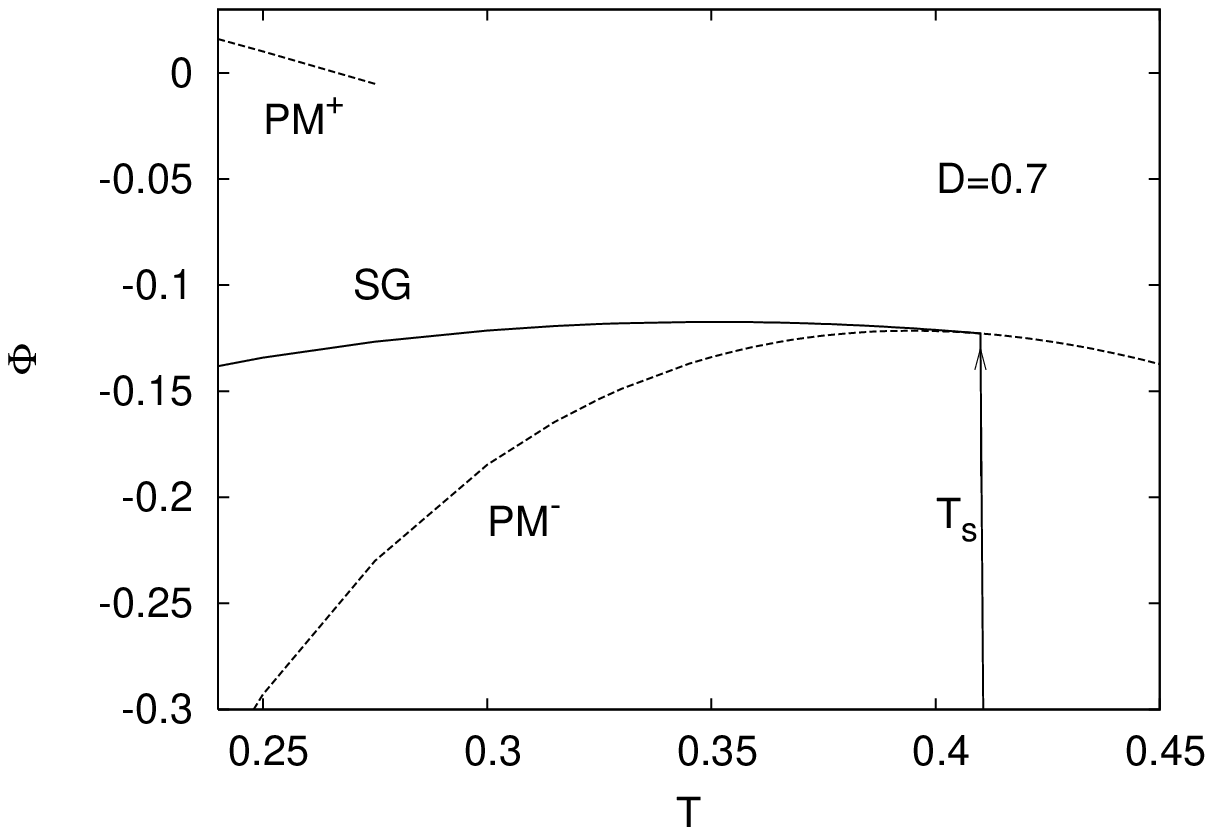}
\includegraphics[width=.47\textwidth]{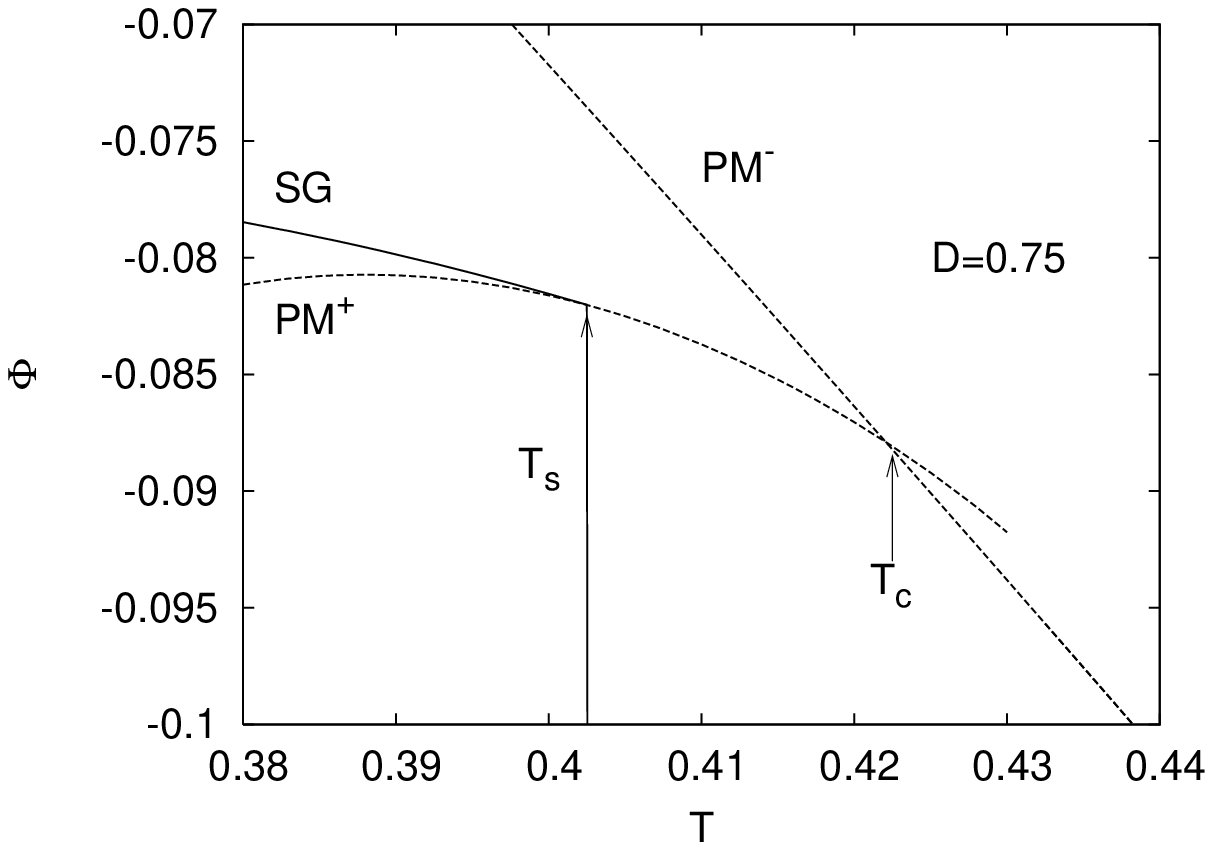}
\includegraphics[width=.47\textwidth]{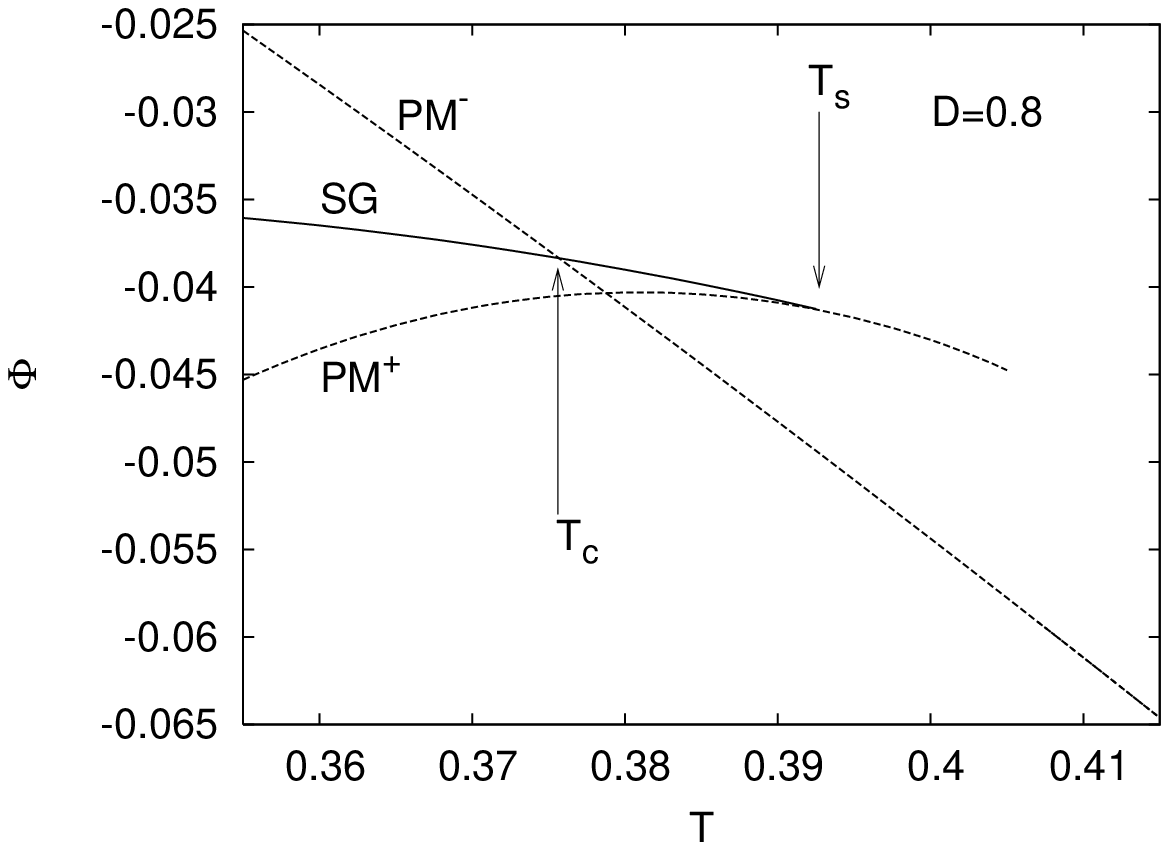}
\includegraphics[width=.47\textwidth]{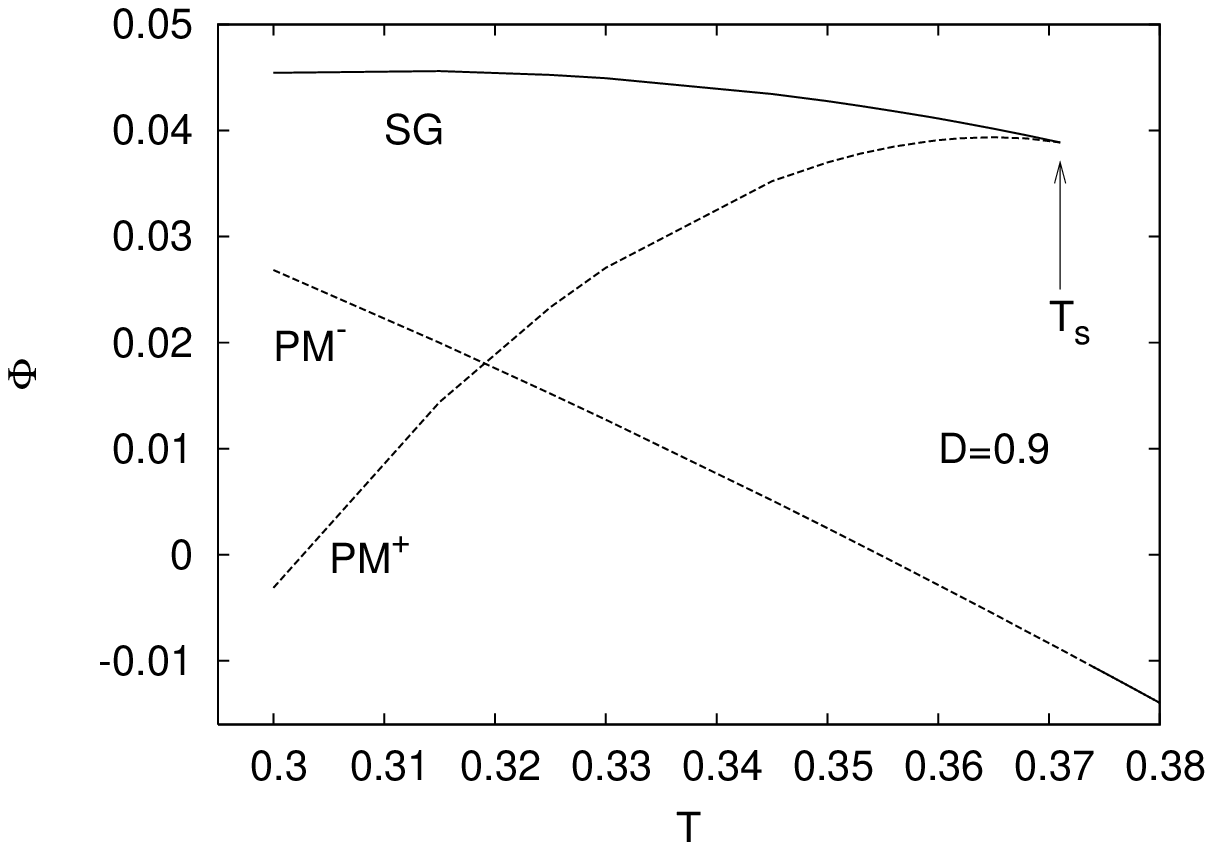}
\caption{Free energy in the SG and PM phases in qualitatively different scenarios. The full curve represents 
$\Phi_{\rm SG}$ in the 1RSB Ansatz , $\Phi_{\mbox{PM}^+}$ and
$\Phi_{\mbox{PM}^-}$ vs. $T$ are plotted as dashed curves.
 $D=0.7,0.75,0.8, 0.9$. 
Top left: at $D=0.7<D_A$, RFOT   between PM$^-$ and SG, at $T_s$. 
Top right: RFOT  between PM$^+$ and SG,
at $T_s$, and TFOT  between PM$^-$ and PM$^+$, at $T_c$, for $D=0.75$
 (larger than $D_A$ and smaller than $D_B$). 
 Bottom left: TFOT  between PM$^-$ and SG, at $T_c$, and RFOT  between 
PM$^+$ and {\em metastable SG}, at $T_s$. 
Bottom right: RFOT  between
  PM$^+$ and metastable SG, at $T_s$.
  }
\label{fig:Phi_compA}
\end{figure}

The SG solution departs from the high-density PM$^+$
phase, i.e., SG and PM$^+$ belong to the same saddle point of
Eq. (\ref{f:FreeEnRSB}), see, e.g., Fig. \ref{fig:Phi_compA}.  For
chemical potential values $D<D_B$ the thermodynamic transition to the
SG is still the RFOT analyzed above. 
In Fig. \ref{fig:Phi_compA} we plot the temperature behavior
of the free energies of the PM$^-$, PM$^+$ and SG phases in three
qualitatively different cases: for $D<D_A$, $D_A<D<D_B$ and $D>D_B$.

\begin{figure}[t!]
\includegraphics[width=.49\textwidth]{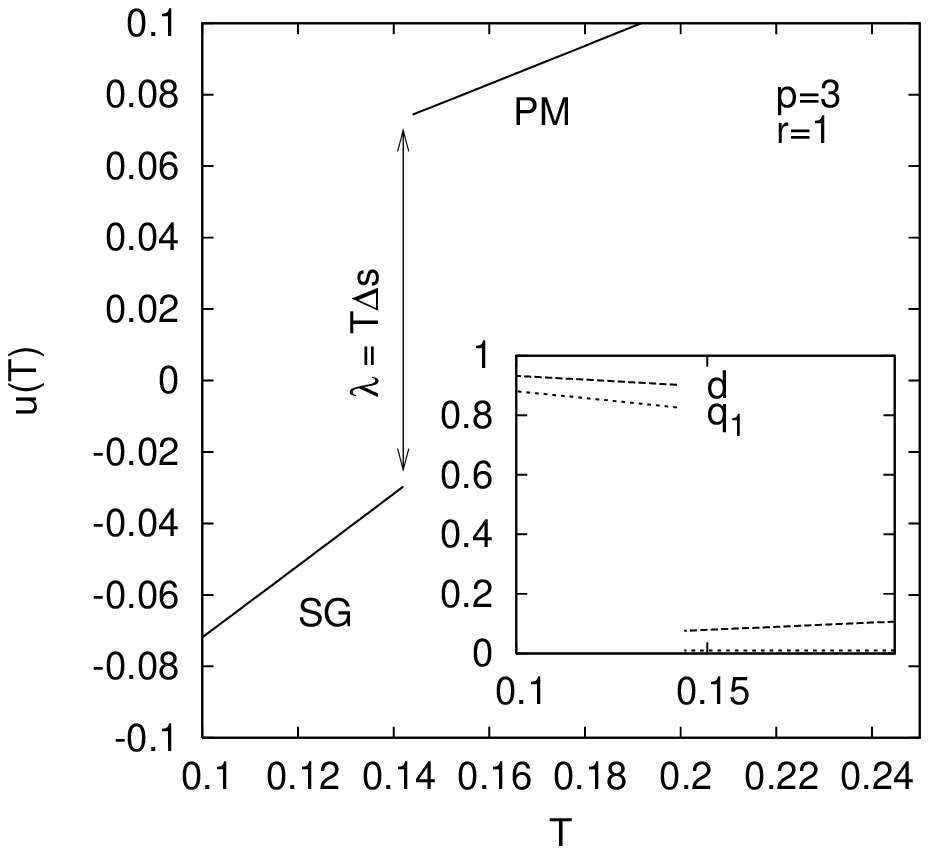}
\includegraphics[width=.49\textwidth]{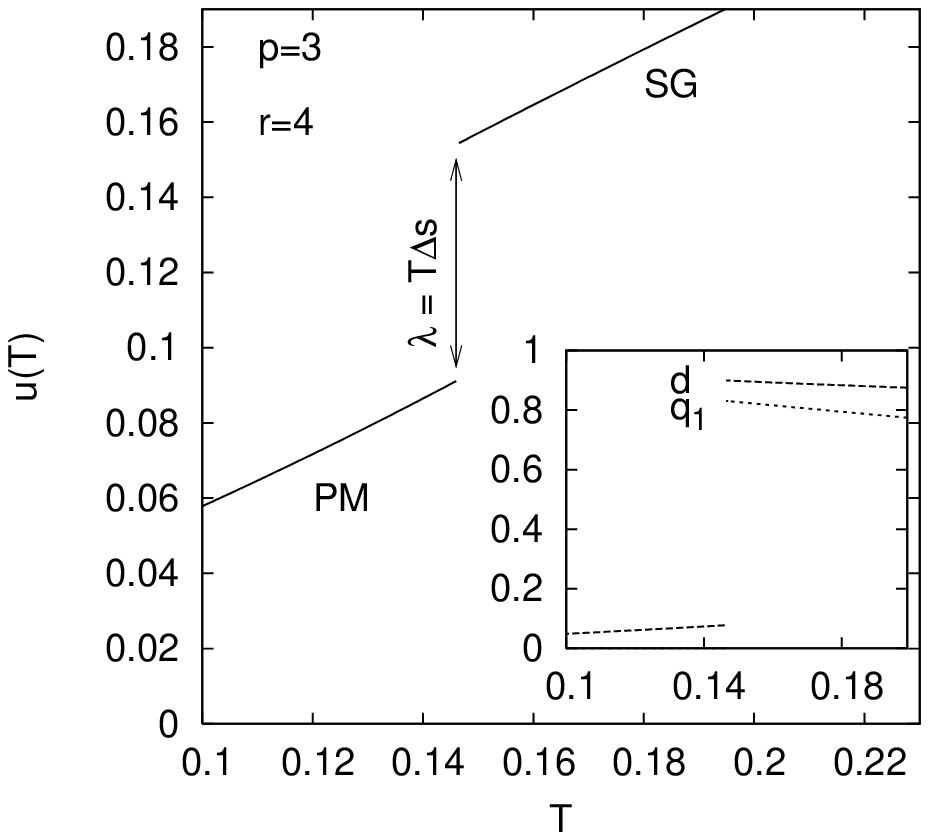}
\caption{Left: Latent heat at the first order transition at $D=1$ for the case $r=1$.  Inset: discontinuity of overlap $q_1$ (dotted line) and density $d$ (dashed line)
across the same transition point at $T_c=0.142$.
Right: Latent heat at $D=1.2$ for the case $r=4$,  in presence of inverse freezing the SG phase transforms into the PM (at lower entropy $\Delta s=s_{\rm SG}-s_{\rm PM}>0$) one as $T$ {\em decreases}, taking heat from the  thermal bath. Inset: the overlap $q_1$ (dotted line) is different from zero above the critical temperature $T_c=0.146$. The density $d$ (dashed line) decreases discontinuously in the low $T$ PM phase.}
\label{fig:latheat}
\end{figure}

Along the $D_c(T)$ transition line, latent heat is taken from the glass
in order to be transformed into a paramagnet, as we show in
 Fig. \ref{fig:latheat}.
 
  Around that line the SG and the PM phases coexist and compete.
Beyond the B point, the RFOT line $T_s(D)$ becomes the SG spinodal
since on the right side of the $D_c(T)$ line the glassy phase is
metastable with respect the PM one. That is, the whole hierarchy of global
and local glassy minima lies above the global thermodynamically dominant PM minimum in the free
energy landscape: $\Phi_{SG}>\Phi_{PM}$. 

This might seem weird in some respect.  Indeed, in RSB theory, e.g.,
in the spherical $p$-spin model, the 1RSB SG solution stems out
of the PM phase, which is RS.  At a given $T_s$ a 1RSB solution
appears with a discontinuous jump in $q_{\rm EA}$ and no discontinuity
in the free energy. As $T<T_s$, the SG free energy is larger than the
PM free energy but it is the one thermodynamically stable because of
the $n-1$, $n-m$ or $m-1$ factors present in the free energy
expression Eq. (\ref{f:FreeEnRSB}) whose sign is inverted in the
$n\to 0$ limit. This implies that looking for a minimal free energy
at finite integer number of replicas corresponds to look for the
maximal of the Parisi free energy. 
In our case, in fact, the SG 1RSB phase stems out of a PM (PM$^+$)
phase, as shown in Fig.  \ref{fig:Phi_compA}.  
However, it competes with {\em another} 
phase PM$^-$, corresponding to a {\em different saddle point} of
Eqs. (\ref{f:FreeEnRSB}), that does not transform into a SG and whose
free energy does not involve overlap terms and, consequently, 
coefficients do not change sign in the replica calculation. 

\begin{figure}[!t]
\center
\includegraphics[height=.26\textheight]{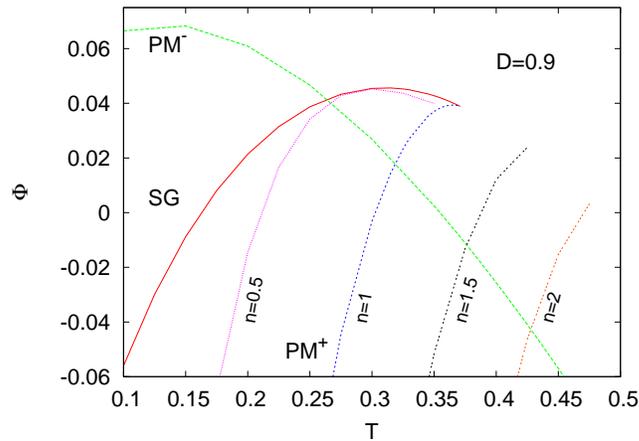}
\caption{Analysis of free energy in temperature at finite $n=0,0.5,1,1,5,2$ in the RS approximation.
The full curve is the SG solution in the zero replica limit.
The dashed curves are finite $n$ SG solutions.
For $n=1$ the SG free energy coincides with the PM$^+$ free energy.
Each one of the SG curves crosses the free energy curve of the PM$^-$ phase signaling a first order phase transition between the PM$^-$ phase, dominant at high $T$ and the SG phase (dominant at low $T$). Since the SG free energy is computed in the RS Ansatz, the exact free energy is a little bit larger and the crossing point at slightly smaller $T$.}
\label{fig:Phi_n_finite}
\end{figure}

 As a homogeneous
selecting rule, we can, thus, apply the minimal free-energy principle
for all competing solutions with $n>1$ and let $n \to 0$ only after
the choice of the dominant stable phase has been done.
To better exemplify this, we plot in Fig. \ref{fig:Phi_n_finite} the
RS approximated SG solution at finite $n$ for different values of $n$
and we compare it with both the free energies of PM$^-$ and PM$^+$.
For $n>1$ the free energy of the SG RS solution crosses the PM$^-$
free energy for $T$ lower than the transition temperature between the two PM phases, signaling
the presence of a transition between the PM$^-$ and the SG.  The
dominant phase is the one of least free energy, i.e., the SG phase.  As
$n$ decreases the critical temperature decreases until, for $n=1$, the
SG free energy turns equal to the PM$^+$ one.  As $n<1$ the SG free
energy shifts further towards small temperature but the thermodynamic order
relationship with the free energy of the PM$^-$ phase is
unchanged. Breaking the replica symmetry on the {\em same} SG phase
leads to a higher free energy and shifts the first order critical
temperature towards smaller $T$.  This does not affect the order
relationship with respect to the {\em other} self-consistent solution, i.e.,  the
PM$^-$ phase.

\subsection{Dynamic-Static inversion}
\begin{figure}[!t]
\center
\includegraphics[height=.26\textheight]{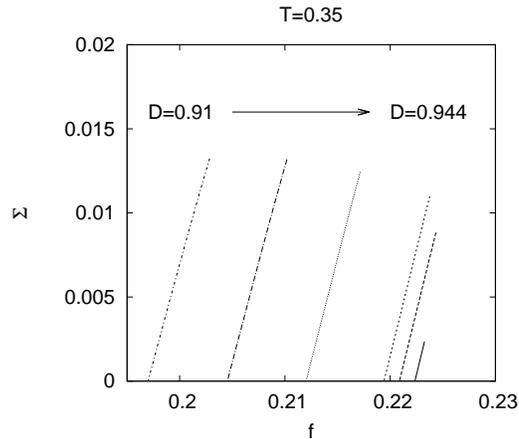}
\caption{
$\Sigma(f)$ for $T=0.35$  at different values of $D$ below the static transition.
}
\label{fig:Sigma_SD_inv}
\end{figure}

As the chemical potential increases, the temperature interval between
$T_s(D)$ and $T_d(D)$ decreases down to zero at  $T_s(D_{I})=T_d(D_I)=T_I=0.3739$, with 
$D_I=0.8905$.  For larger $D$ the dynamic and static lines
invert their position in $T$.  Coming from high temperature, this
means that the static line ($m=1$ in Eq. (\ref{f:m_static})) will be
met before the dynamic one ($m=1$ in Eq. (\ref{MAX_COMPL})). According
to the behavior of the complexity this implies that the lowest
equilibrium glassy states are dynamically accessible at the static
transition
\footnote{The RFOT static transition line is actually a spinodal,
since the thermodynamic dominant phase is PM beyond the $D_c(T)$
line.}  and that excited glassy states only develop as $T$ is {\em
decreased}, as displayed in Fig.  \ref{fig:Sigma_SD_inv}.

Actually, it can be seen from the complete stability analysis, cf . \ref{app:1RSB_stab},
 that for $D>D_{E_d}= 0.8737$,
 slightly smaller than $D_I$, the SG dynamic solution
spanned along the $m=1$-line
turns out to display a couple of complex conjugated eigenvalues with {\em negative real part}.
This implies that for $D\geq D_E$ the first thermodynamically stable SG solution occurs with the static parameter $m<1$ and zero complexity for the excited states.
The static-dynamic inversion then occurs, self-consistently, for values of $m<1$, at $D_{I_{\rm stab}}=0.877$, $T_{I_{\rm stab}}=0.377$.

\subsection{$m$-lines inversion, compressibility and replica stability}
\begin{figure}
\center
\includegraphics[width=.48\textwidth]{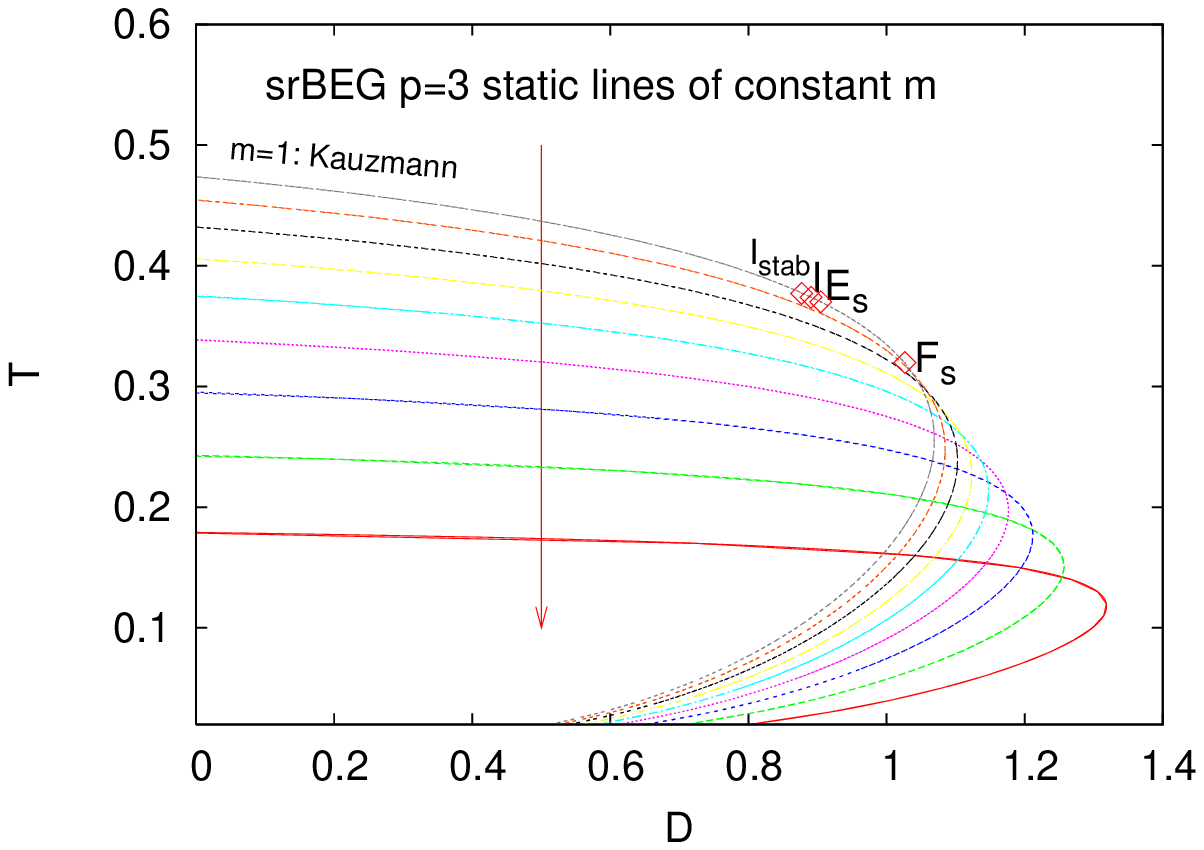}
\includegraphics[width=.48\textwidth]{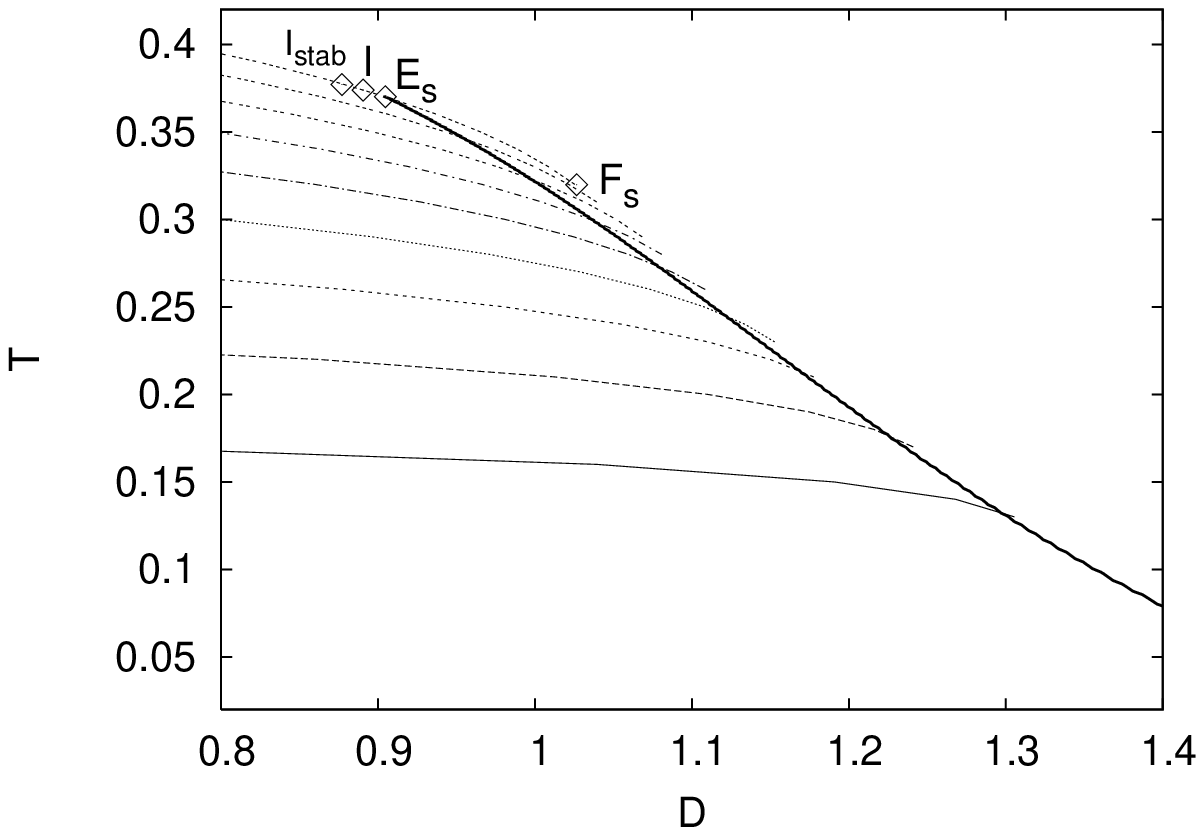}
\caption{$T(D)$ lines of constant $m$ (from static equations). As $D$ is large enough they bend and cross each other. Left: complete behavior of $m$-lines. Right: only curves along which the compressibility is non-negative are plotted. An equivalent situation occurs for the dynamic $m$-lines.
The thick continuous curve is the line where complex stability eigenvalues acquire negative real part: solutions on the right hand side of the stability line are rejected.}
\label{f:minversion}
\end{figure}

As one can see from Fig. \ref{f:minversion}, solving equations
(\ref{f:sceq1_p_RSB}), (\ref{f:sceq2_p_RSB}) and
Eq. (\ref{f:m_static}) [else (\ref{MAX_COMPL})], and following
solutions at fixed $m$ in the $D,T$ plane, each $T_m(D)$ curve
("$m$-line") develops a maximum and $T_m(D)$ lines at different $m$
values cross.  This leads to an ambiguity: there is a region in the
phase diagram in which in each point two different solution of
Eqs. (\ref{f:sceq1_p_RSB}), (\ref{f:sceq2_p_RSB}),(\ref{f:m_static})
occur.  A criterium is, then, required to select the correct (metastable)
phase in each point and, furthermore, a way to mark the spinodal of
the SG. The solution comes from the positiveness of the
compressibility $\kappa$ and from the complete analysis of the stability of the
1RSB solution in the replica space.
Moreover, as we already mention, the $m$-line inversion comply with the fact  that
the spinodal SG lines (both dynamic and static) will cease to be a line
of constant $m=1$.

\subsection{Compressibility}
\label{ss:compress}
As a thermodynamic quantity the compressibility is usually defined as: 
\begin{equation}
k = -\left. \frac{\partial \log V}{ \partial p}\right|_T
\end{equation}
In our model, where the role of the pressure is played by the chemical
potential $D$ and the inverse specific volume is the density of
filled-in states $d$, the compressibility reads:
\begin{equation}
k=  d \left. \frac{\partial d^{-1}}{\partial D}\right|_T = - \frac{1}{d} \left. \frac{\partial d}{\partial D} \right|_T = 
  -\frac{1}{d} \left. \frac{\partial d}{\partial m} \right|_T  \left. \frac{\partial m}{\partial D} \right|_T 
\label{f:compressibility}
\end{equation}
where the last equality comes from the Fubini's implicit function
theorem. Obviously, solutions with negative compressibility are
unphysical and have to be discarded.  At fixed $T$ the value $m=\hat m$ yielding the zero point of the compressibility is given by \begin{equation}
 \left. \left( \left. \frac{\partial m}{\partial D} \right|_T \right) \right|_{m=\hat m}=0
\end{equation} 
using, e.g., Eq. (\ref{f:sceq2_p_RSB}) for the $D(m)$ dependence.
 Beyond $\hat m(T)$
  the compressibility turns negative and that solution will
 be rejected.  This kind of "selection rule" has to be used for both
 the static and the dynamic \textit{m-lines}. 
 
 The positiveness of $\kappa$ is certainly a necessary condition for thermodynamic stability. On top of that we anticipate that, from the replica stability analysis, we found a small region of the phase diagram where $\kappa>0$ but the real part of a couple of complex-conjugated eigenvalues among $\Lambda_{2,1}$, or among  $\Lambda_{3,1}$
  of the Hessian in the replica space for the 1RSB Ansatz, cf. Eqs. (\ref{Hess:S21}), (\ref{Hess:S31}),
   is negative (see also the right panel of  Fig. \ref{f:minversion}). The spinodal lines plotted in Figs. \ref{fig:PhDi_p_TD}, \ref{fig:PhDi_p_TD_r4} are drawn taking the latter observation 
into account. See also the right panel of Fig. \ref{f:minversion}

\begin{figure}
\center
\includegraphics[width=.48\textwidth]{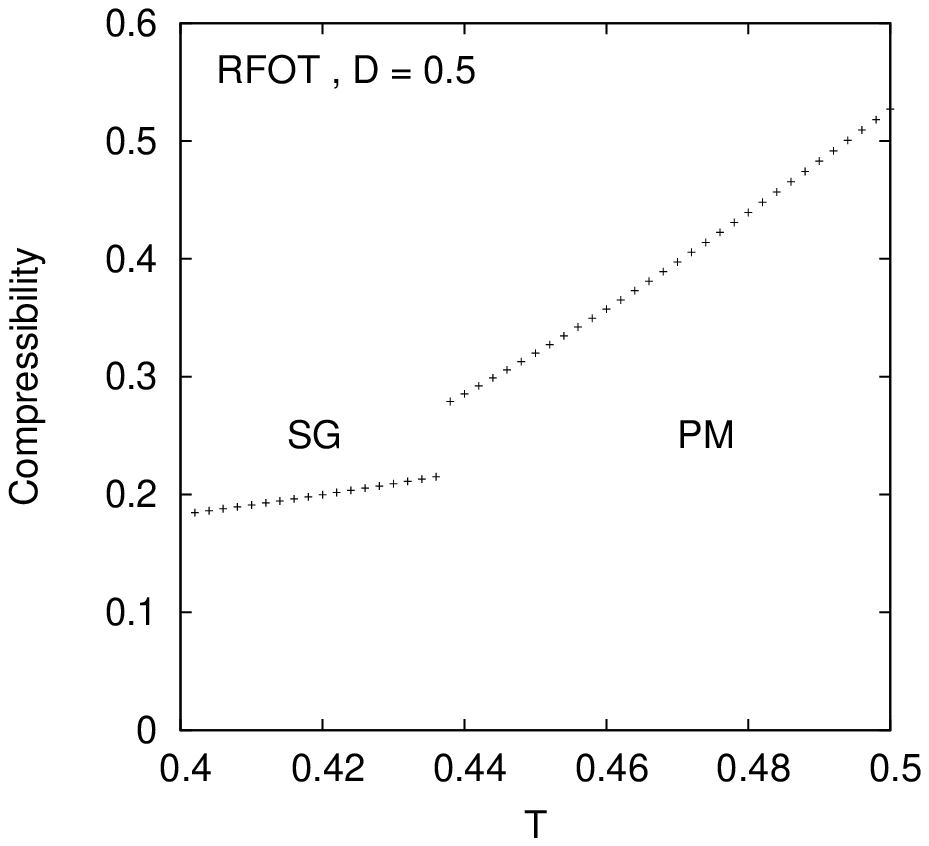}
\includegraphics[width=.48\textwidth]{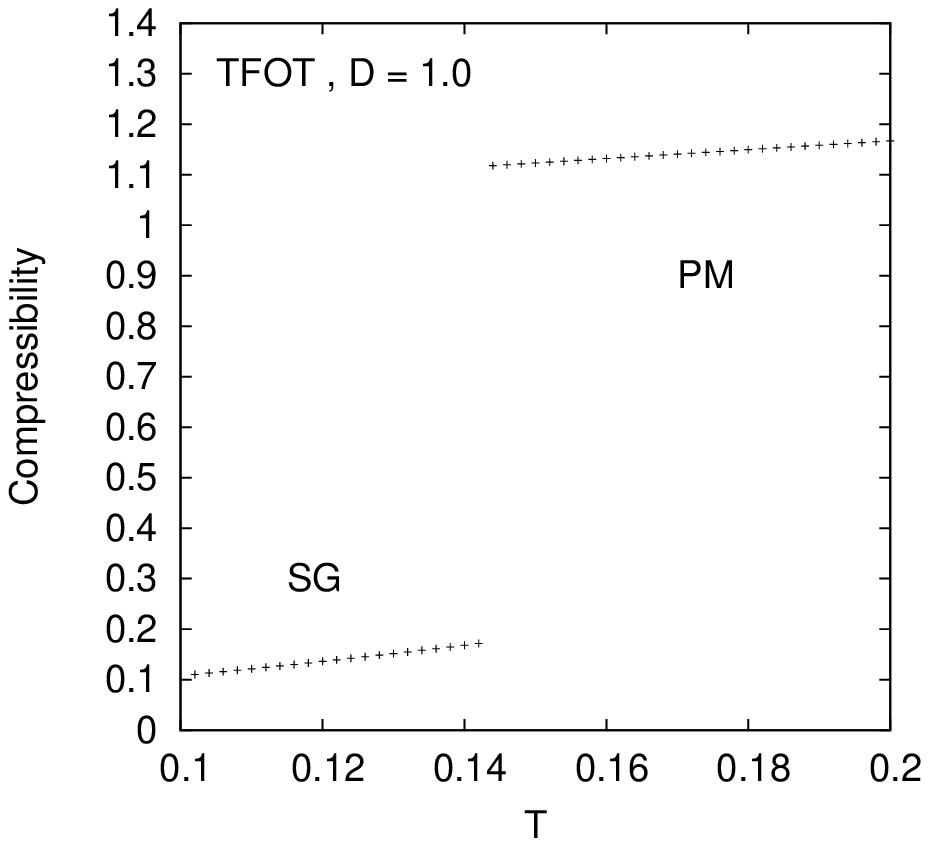}
\caption{Compressibility vs, $T$  at $D=0.5<D_A$, in the RFOT regime (left) and at $D=1>D_B$ in the TFOT regime (right).}
\label{f:compress}
\end{figure}

The previous discussion implies that the dynamic SG spinodal is the
 $m=1$-line on the left side of the point $F_d$, $(D_{F_d}=0.98,T_{F_d}=0.33)$ and a line
 of decreasing $m$ on the right side.  Also the static SG spinodal
 undergo the same change at the point $F_s$ $(D_{F_s}=1.03,T_{F_s}=0.32)$.  We
 remark, however, that this change in behavior arises for $D>D_{I_{\rm stab}}$, where the $m=1$-line cease to be stable in the replica space and $D>D_I$, after that the
 static-dynamic inversion occurs.

\subsection{Stability of 1RSB solution}
\label{sec:stab}
In order to determine the above phase diagram it has been necessary to
study the stability of the 1RSB solution in the $T,D$ plane.  In
 \ref{app:1RSB_stab} we show the details of our analysis.  The
phenomenology looks similar to the RS case with both real and complex
eigenvalues. The stability of the SG solution is ruled by
$\Lambda_{1,1}^{(3)}$, the lowest real eigenvalue of the matrix $H_{11}$, cf. Eq.
(\ref{H_11}).  Evaluated on the static SG solution it is always
positive, thus, according to our analysis, the static SG 1RSB solution
is always stable. If, however, the static Eq. (\ref{f:m_static}) is
not imposed and the parameter $m$ is, thus, left undetermined, the positiveness
of $\Lambda_{1,1}^{(3)}$ is equivalent to the condition:
\begin{equation}
 \eta_1 > (p-1) \eta_0
\end{equation}
and the equality $\Lambda_{1,1}^{(3)}=0$ turns out to coincide with
the dynamical condition Eq. (\ref{MAX_COMPL}). The other real
eigenvalues are positive in the whole phase diagram.

In the region of phase coexistence, near the SG spinodal, the stability analysis actually revelas  more subtle features. 
Indeed, as in the $p=2$ case, cf. Sec. \ref{s:p=2}, 
 complex eigenvalues are present and some of them develop a
negative real part. In particular, solutions with negative compressibility always yield an
imaginary eigenvalues with negative real part. Moreover, in a small region of the $T$,$D$ plane solutions with 
 $\kappa\gtrsim 0$ also display complex eigenvalues, $\Lambda_{2,1}$ or $\Lambda_{3,1}$, cf. Eqs.
 (\ref{Hess:S21}), (\ref{Hess:S31}) with negative real part.
As anticipated,  in determining the
phase diagram we have thus rejected solutions with negative replica
eigenvalues, including all those with negative compressibility.

\begin{figure}[!t]
\center
\includegraphics[height = .26\textheight]{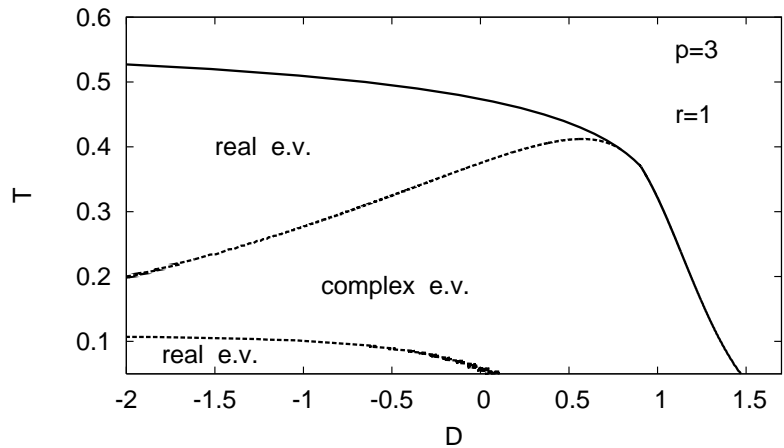}
\caption{Region of complex  eigenvalues  in the $T,D$ plane for the $p=3$ model.}
\label{fig:PhDi_eigenvalues}
\end{figure}

\section{Conclusions}
We have been analyzing, at the level of mean-field theory,
 a family of models mixing the properties of (i)  Blume-Capel models with quenched disorder and
 (ii) spin-glass models with many-body ($p>2$) interactions.
 The first ones are
 known to reproduce 
the inverse freezing occurring, e.g.,  in complex macromolecular compounds in solution \cite{P4MP1,aCD,MetCel},
 and display first order thermodynamic phase transitions even in presence of strong disorder,
The latter yield most of the basic features  of the standard folklore of glassy systems and are often refereed to as mean-field glasses.
Our aim has been to investigate to behavior of dynamic arrest, displayed in mean-field glasses, as a
system undergoes an inverse thermodynamic transition.

We have been studying the problem by means of replica theory for the two qualitatively distinct cases with $p=2$ and  $p>2$, both 
looking at the thermodynamic and at the dynamic properties. The dynamics has been tackled by means of the analysis of the complexity of the
free energy landscape. The model approximation used is to pass from discrete spin-$1$ variables to spherical ones (i.e., continuous variables with an overall "spherical" constraint) keeping the original probabilistic relationship between states of the discrete spin: $r\equiv$ $[\#$ of times $s=1]/$  $[\#$ of times $s=0]$. This allows for thoroughly analytical computation of complicated physical quantities. 
We have been mainly focusing on the case $r=1$, i.e., the spherical formulation of the original BC model but we have considered, as well, model cases with different $r$ values. Indeed, in those cases where 
the spherical analogue of the original spin-$1$ system ($r=1$) do not show inverse freezing, a simple variant of the model for larger $r$ is suitable to describe such  phenomenon. 

The external parameters driving possible transitions are the temperature and a chemical potential for the empty states of the spin-$1$ variables,
the so-called "crystal field" $D$.  For $D=-\infty$ one has no holes and the limit of the model is the spherical spin-glass. In that limit, in the $p=2$ case  \cite{Kosterlitz76} the spin-glass phase is described by a replica symmetric Ansatz and the transition turns out to be always continuous, whereas, as $p>2$ the RS solution is unstable and the right physics is described by a one step Replica Symmetry Breaking phase \cite{Crisanti92}. On top of it the transition is not continuous (at zero external magnetic field) in the order parameter and it is often refereed to as random first order.

In the present study we find that, for $p=2$, the situation is not very different for the original spherical spin-glass, with a spin-glass phase stable under the RS Ansatz. The only addition is the presence of a 
transition between competing paramagnetic  phases at high $T$ ($D$) and the occurrence of inverse freezing for large enough chemical potential $D$. 
For $p>2$ the situation is, instead, drastically modified.
Besides the standard RFOT, we find that, for large enough $D$, a first order thermodynamic phase transition between the spin-glass and the paramagnetic phase takes over, with latent heat and jumps in density and overlap order parameter. The spin-glass becomes metastable  with respect to the paramagnet and the RFOT line plays the role of the spin-glass spinodal.
Moreover, analyzing in detail the structure of the solutions in the phase diagram and their stability properties, we find that the spin-glass phase can be approached, in given regions, without incurring in dynamic arrest, unlike in the standard $p$-spin and that the first spin-glass solution display values of replica symmetry breaking parameter $m$ less than one.
The study of the complexity functional clarify how this observation amounts to a scenario in which, in a cooling experiment, lowest
glassy minima of the free energy landscape develop first and excited metastable glassy states only
arise as
temperature is further lowered. This provides a model case where actual dynamics might be followed
along given paths of the phase diagram into the deep glassy phase without being stuck in the threshold states.

\section*{Acknowledgements}
The authors thank Andrea Crisanti and Giorgio Parisi for stimulating discussions. 
The research leading to these results has received
 funding from the Italian Ministry of Education,
 University and Research under the Basic Research
 Investigation Fund (FIRB/2008)  program/CINECA grant code RBFR08M3P4.

\appendix
\section{Hessian of the spherical $p$-spin Blume-Capel model}
\setcounter{section}{1}
\label{app:Hessian}

In order to verify the stability of a saddle-point calculation, the
positiveness of the fluctuation matrix evaluated on the stationary
solution has to be checked. This Hessian matrix is constructed as the
second order variation of the free-energy potential
(\ref{f:FreeEnMatr}) with respect to the three overlap matrices
(\ref{f:Q}) and, indeed, depends on two couples of indexes:
\begin{equation}
 H(G)^{O,O'}_{ab;cd} = \frac{\partial^2 G}{\partial O_{ab} \partial O'_{cd}}, \quad O,O' = Q,S,T,
 \label{Hess}
\end{equation}

In detail the elements read

\begin{eqnarray*}
 \frac{\partial G^2}{\partial Q_{ab} \partial Q_{cd}} &=&-B_{ab} \delta_{a,c}\delta_{bd} + \frac{1}{8} \left[
 A_{a,c} (TAT)_{bd} + 2(TA)_{a,c} (AT)_{bd} +\right. \\
&&\left.+  (TAT)_{a,c} A_{bd} \right] \\
\frac{\partial G^2}{\partial Q_{ab} \partial T_{cd}} &=& - B_{ab} \delta_{a,c}\delta_{bd} + \frac{1}{8} \left[A_{a,c} (TAQ)_{bd} + (QA)_{ac} (AT)_{bd} + \right.\\
&&\left.  + (AT)_{ac} (AQ)_{bd} +(TAQ)_{ac} A_{bd} \right] - \frac{1}{4}\left( A_{bd} \delta_{ac}+ A_{ac} \delta_{bd}\right)\\
\frac{\partial G^2}{\partial Q_{ab} \partial R_{cd}} &=&  - 2B_{ab} \delta_{ac}\delta_{bd} - \frac{1}{4} \left[A_{ac} (TAR)_{bd} + (RA)_{ac} (AT)_{bd} \right.  \\ 
&&+ \left. (AT)_{ac} (AR)_{bd} +(TAR)_{ac} A_{bd} \right]
\end{eqnarray*}
\begin{eqnarray*}
\frac{\partial G^2}{\partial T_{ab} \partial Q_{cd}}  &=& - B_{ab} \delta_{ac}\delta_{bd} + \frac{1}{8} \left[A_{ac} (TAQ)_{bd} + (QA)_{ac} (AT)_{bd} + \right. \\ 
&&+ \left. (AT)_{ac} (AQ)_{bd} +(TAQ)_{ac} A_{bd} \right] - \frac{1}{4}\left( A_{bd} \delta_{ac}+ A_{ac} \delta_{bd}\right)\\
 \frac{\partial G^2}{\partial T_{ab} \partial T_{cd}} &=& -B_{ab} \delta_{ac}\delta_{bd} + \frac{1}{8} \left[A_{ac} (QAQ)_{bd} + 2(QA)_{ac} (QA)_{bd} + \right. \\
&+& \left. (QAQ)_{ac} A_{bd} \right] \\
\frac{\partial G^2}{\partial T_{ab} \partial R_{cd}}&=& - 2B_{ab} \delta_{ac}\delta_{bd} - \frac{1}{4} \left[A_{ac} (QAR)_{bd} + (RA)_{ac} (AQ)_{bd} \right.  \\ 
&&+ \left. (AQ)_{ac} (AR)_{bd} +(QAR)_{ac} A_{bd} \right] 
\end{eqnarray*}
\begin{eqnarray*}
\frac{\partial G^2}{\partial R_{ab} \partial Q_{cd}} &=&  - 2B_{ab} \delta_{ac}\delta_{bd} - \frac{1}{4} \left[A_{ac} (TAR)_{bd} + (RA)_{ac} (AT)_{bd} \right.  \\ 
&&+ \left. (AT)_{ac} (AR)_{bd} +(TAR)_{ac} A_{bd} \right] \\
 \frac{\partial G^2}{\partial R_{ab} \partial T_{cd}} &=& - 2B_{ab} \delta_{ac}\delta_{bd} - \frac{1}{4} \left[A_{ac} (QAR)_{bd} + (RA)_{ac} (AQ)_{bd} \right.  \\ 
&&+ \left. (AQ)_{ac} (AR)_{bd} +(QAR)_{ac} A_{bd} \right] \\
\frac{\partial G^2}{\partial R_{ab} \partial R_{cd}} &=& - 4B_{ab} \delta_{ac}\delta_{bd} + \frac{1}{2} \left[A_{ac} (RAR)_{bd} + 2(RA)_{ac} (AR)_{bd} + \right. \\
&&+ \left. (RAR)_{ac} A_{bd} \right] + \frac{1}{2}\left( A_{bd} \delta_{ac}+ A_{ac} \delta_{bd}\right) \\
\end{eqnarray*}
where:
\begin{eqnarray}
A_{ab} &=&(QT-R^2)_{ab} \\
B_{ab} &= &\frac{\beta^2 p (p-1)}{2^6} \left( \frac{Q_{ab}+T_{ab} + 2R_{ab}}{4}\right)^{p-2} 
\label{AeB}
\end{eqnarray}

To check its positiveness of the Hessian, one needs to compute, and determine the
sign of all the eigenvalues of its action on fluctuation matrices
($\delta o$'s):
\begin{eqnarray}
\Lambda ~ \delta o_{ab}.&=& \sum_{cd} \left( \frac{\partial^2
G}{\partial O_{ab} \partial Q_{cd}} \delta q_{cd} + \frac{\partial^2
G}{\partial O_{ab} \partial T_{cd}} \delta t_{cd} + \frac{\partial^2
G}{\partial O_{ab} \partial R_{cd}} \delta r_{cd} \right) = \\ &=& -
B_{ab} \delta o_{ab} + \frac{1}{2} \sum_{cd} \left( \sum_{O'} \frac{
\partial^2 }{\partial O_{ab} \partial O'_{cd}} \log | A | \right)
\delta o_{cd}.
\label{Hessdiag}
\end{eqnarray}

In order to solve Eq. (\ref{Hessdiag}), a set of $ \frac{3n^2 -n}{2}$
coupled equations, one needs to find a way to project the system on
proper subspaces, that allows to translate the calculation to an usual
eigenvalues problem.

\section{Stability of the RS solution}
\setcounter{section}{2}
\label{app:RS_stab}

In order to understand the appearance of complex eigenvalues, we start
the stability analysis retaining $n$ finite. In the RS ansatz the
matix (\ref{AeB}) becomes:
\begin{eqnarray}
A_{ab}^{-1} &=& \frac{1}{4\theta \eta}\delta_{ab} 
- \frac{q}{4\theta \eta \eta_1} \equiv x \delta_{ab} + w \\
&&x\equiv \frac{1}{4\theta\eta}
\\
&&w\equiv -\frac{1}{4\theta\eta\eta_1}
\\
B_{a\neq b} &=& \frac{1}{32}\phi'(q)=
 \frac{\beta^2 p (p-1)}{2^6} q^{p-2} \equiv B_q,  \\
 B_{a=b}&=& \frac{1}{32}\phi'(d)=
 \frac{\beta^2 p (p-1)}{2^6} d^{p-2} \equiv B_d
\end{eqnarray}
where $\theta = 1-d$, $\eta = d-q$ and $\eta_1 = d - (1-n) q$, with
$\eta_1 \to \eta$ if $n \to 0$. The parameter $x$ is always positive.
For later convenience, we define:
\begin{eqnarray}
 D_{++}&\equiv& x^2 (\eta + \theta)^2 ;\qquad  \quad \, \, \,
 H_{++}\equiv
 x w [ \eta^2 + 4  \theta^2+ \eta (4\theta- \eta_1)];
 \label{f:DH++}
 \\
 D_{+-} &\equiv& (\eta+ \theta)(\eta-\theta);
 \qquad H_{+-} \equiv x w ( \eta^2 - 4  \theta^2- \eta \eta_1);
  \label{f:DH+-}
\\
 D_{--}& \equiv& (\eta-\theta)^2 
 ;\qquad  \qquad \, \, \, 
H_{--} \equiv x w ( \eta^2 + 4  \theta^2- \eta (2 \theta+ \eta_1)
 \label{f:DH--}
\end{eqnarray}
\subsection*{Projection on subspace $S_1$}
Fluctuations that involve the overlap of one replica with the others,
are selected by projecting the Hessian (\ref{Hess}) in the subspace
$S_1$ defined by: $$\delta r_{a,a}=0 , \quad \sum_{\gamma} \delta
q_{a,\gamma}=0$$ with dimension $d(S_1)= d_{tot} - 4 n = \frac{3 n
(n-3)}{2}$. This is called replicon subspace and in $S_1$ the Hessian matrix reduces to:
\begin{equation}
\begin{array}{c}
 H_1=
\left(
\begin{array}{c c c}
\frac{1}{2} D_{++} - B_q  & \frac{1}{2}D_{++}-\frac{x}{2} - B_q  &  -D_{+-} -2B_q  \\
 \frac{1}{2} D_{++}- \frac{x}{2} -B_q & \frac{1}{2}D_{++} - B_q  & -D_{+-} -2B_q  \\
 - D_{+-} - 2B_q & -D_{+-} -2 B_q &  2 D_{--} +x -4B_q \\
\end{array} \right)
\end{array}
\label{H1p2}
\end{equation}
displaying no explicit $n$ dependence. The characteristic
polynomial, evaluated on the SG solution, reads:
\begin{eqnarray}
\label{PLambda_1}
 P[H_1] &=& \left(\Lambda-\frac{x}{2}\right)(\Lambda^2 - b_1 \Lambda + c_1) \\
 b_1 &=& \frac{3 \left(\eta^2-(p-2) \theta^2 \right)}{16 \eta^2 \theta^2} \\
 c_1 &=& -\frac{p-2}{8 \eta^2 \theta^2}
\end{eqnarray}

\subsubsection*{Interaction with $p=2$\\}
For $p=2$ $c_1=0$, $b_1>0$: two eigenvalues will be positive  and one will be zero. 
\begin{eqnarray}
 \Lambda_1^{(1)} &=& 0
 \label{f:Replicon_p2}\\
 \Lambda_1^{(2)} &=&  \frac{3}{16\theta^2} > 0\\
 \Lambda_1^{(3)} &=&  \frac{1}{8\theta\eta} > 0
\end{eqnarray}
We have marginal stability of the RS-SG solution.

\subsubsection*{Interaction with $p>2$ and replica symmetry breaking\\}

 If $p>2$, instead, $c_1<0$, leading to a negative zero of the polynomial
and causes the instability of the RS-SG solution. 
As $p>2$ one has to resort to a more complicated Ansatz, the 1RSB scheme of computation, whose stability will be analyzed in the following appendix.
We now continue the analysis of the RS fluctuations in other subspaces.

\subsection*{Projection on subspace $S_2$}
Fluctuations of the cluster of replicas are selected by projecting in
$S_2$, defined as: $$\sum_{\gamma} \delta r_{\gamma,\gamma}=0 , \quad
\sum_{\gamma,\delta} \delta q_{\gamma,\delta}=0$$ with $d(S_2)= 4 n -
4$. In $S_2$ the Hessian reduces to:
\begin{eqnarray}
\hspace*{-2.5cm}\begin{array}{c}
H_2 = H_1'+ 
\left(
\begin{array}{c c c c}
\frac{n-2}{8} H_{++} & \frac{n-2}{8}(H_{++}-2 w) &
-\frac{n-2}{4}H_{+-} & -\frac{n-2}{4}H_{+-}\\ \frac{n-2}{8}
(H_{++}-2w) & \frac{n-2}{8}H_{++} & -\frac{n-2}{4}H_{+-} &
-\frac{n-2}{4}H_{+-}\\ -\frac{n-2}{4}H_{+-} & -\frac{n-2}{4}H_{+-}
& \frac{n-2}{2}(H_{--}+w) & \frac{n-2}{2}(H_{--} +w)\\
-\frac{1}{2} H_{+-} & -\frac{1}{2} H_{+-} & H_{--} +w & H_{--}
+w\\
\end{array} \right)
\end{array}
\nn
\\
\label{H_2}
\end{eqnarray}
where:
\begin{equation}
H_1':=
\left(
\begin{array}{c|c}
 H_1  & 0 \\
\hline
 0 & \Lambda_4 \\
\end{array}
\right)
\end{equation}
with 
\begin{equation}
\label{f:Lambda4}
\Lambda_2^{(4)}= - 4 B_d + x + 2 D_{--}
\end{equation}
 For generic $n$, apart from $n = 4$, $H_2$ is not symmetric, so the matrix can have complex eigenvalues. 
  The
 characteristic polynomial of the $H_2$ matrix, evaluated on the SG
 solution, reads:
\begin{eqnarray}
&&\hspace*{-2cm} P[H_2](\Lambda)=\left(-\Lambda + \frac{x}{2}+ \frac{n-2}{4}w
 \right)\left[- \Lambda^3 + a_2(n) \Lambda^2 - b_2(n) \Lambda +
 c_2(n) \right]
\nn
\\
&&\label{Ph2}
\end{eqnarray}
In the limit $n\to 0$, the coefficients are:
\begin{equation}
\begin{array}{c c c}
 a_2 &=& - 2 w \eta^2 x^3 \\
 b_2 &=&  - w x + 6 x^4 \eta^4\\
 c_2 &=&  5 x^2 \eta^2 - 2 w x \theta^2\\
\end{array}
\end{equation}

\subsubsection*{Interaction with $p=2$\\}
For the case $p=2$,
a numerical study ensures that all the solutions of the characteristic
polynomial are real positive or complex conjugated with a positive real part. The region of
the $(D,T)$ plane where the polynomial has complex zeros (shown in Fig
\ref{fig:imaginary_ev}) can be determined imposing:
\begin{equation}
 \frac{u^2}{4}+\frac{v^3}{27}>0
\end{equation}
where (in terms of the polynomial coefficients):
\begin{equation}
 \begin{array}{c c c}
u &=& \frac{2 a_2^3}{27} + \frac{a_2 b_2}{3} + c_2 \\
v &=& - b_2 - \frac{a_2^2}{3}
\end{array}
\end{equation}

\subsubsection*{Stability of the  paramagnetic phase\\}
In the paramagnetic phase $w\propto q=0$, implying $H_{\pm\pm}=0$, cf. Eqs. (\ref{f:DH++})-
(\ref{f:DH--})
and the Hessian simplifies as $H_2 = H_1'$.
Its
characteristic polynomial is:
\begin{equation}
 P[H_2]= (\Lambda-\Lambda_4)\left(\Lambda-\frac{x}{2}\right)(-\Lambda^2 - b_2 \Lambda + c_2)
\end{equation}
where now, cf. Eq. (\ref{PLambda_1}):
\begin{eqnarray}
 b_2 &= \frac{3}{16}\left(\frac{2d^2-2d+1}{(1-d)^2 d^2}-\beta^2 \delta_p^2 \right) \\
 c_2 &=\frac{1-\beta^2d^2 \delta_p^2 }{32(1-d)^2d^2}
\end{eqnarray}
Its zeros are:
\begin{eqnarray}
 \Lambda_1^{(2)}&=&\frac{x}{2}\\
 \Lambda_2^{(2)}&=&\frac{-b_2 + \sqrt{b_2^2 - 4 c_2}}{2}\\
  \Lambda_3^{(2)}&=&\frac{-b_2 - \sqrt{b_2^2 - 4 c_2}}{2}\\
\Lambda_4^{(2)} &=& \frac{1}{8} \left( \frac{1-2d+2 d^2 }{d^2(1-d)^2} - \frac{p(p-1)\beta^2}{2}\right) 
\end{eqnarray}

The first one is strictly positive for any $p$. 

The second one is always positive for $p>2$ and
ensures the stability of the PM solution in the whole phase-space. For
$p=2$, instead, $\Lambda_2^{PM}$ marks the PM solution as unstable in
the region where a SG solution exists and permits to discard an
unphysical solution preventing a PM-PM first order
transition. $\Lambda_3^{PM} > \Lambda_2^{PM}$ and, indeed, it is
irrilevant.

The fourth eigenvalue, equal to the derivative of
Eq. (\ref{f:sceq2_p_RSB}) with $q=0$, allows to discard another 
unphysical solution among the PM solutions.



\subsection*{Projection on subspace $S_3$}
In the subspace $S_3= (S_1 \cup S_2)^{\perp}$ ($d(S_3)=4$) the Hessian
reduces to a matrix $H_3 = H_2 + O(n)$; indeed there are no new
eigenvalues. \\

\section{Stability of the 1RS solution}
\setcounter{section}{3}
\label{app:1RSB_stab}

The paramagnetic solution is not affected by the Ansatz choice; its
stablity analysis is indeed the same as the one in
\ref{app:RS_stab}. Moreover for $p=2$ is not necessary to break
the replica symmetry, so in this section we choose $p>3$ without loss
of generality. These observations allow us to work directly with Eq.'s
(\ref{f:sceq1_p_RSB},\ref{f:sceq2_p_RSB}) imposed. Instead, in order
to study also dynamical solution Eq. (\ref{f:m_static}) is not
imposed. \\ In the 1RSB ansatz the matix (\ref{AeB}) becomes:
\begin{eqnarray}
A_{ab}^{-1} &=& \frac{\delta_{ab}}{4\theta \eta_0} - \frac{\eta_1-\eta_0}{2 m \theta \eta_0 \eta_1} \equiv x \delta_{ab} + w \\
&&x\equiv \frac{1}{4\theta\eta_0}\\
&&w\equiv -\frac{\eta_1-\eta_0}{2m\theta\eta_0\eta_1}\\
 \\
B_{a\neq b} &=& \frac{1}{32}\phi'(q_1)\epsilon_{ab}=
 \frac{\beta^2 p (p-1)}{64} q_1^{p-2}\epsilon_{ab} \equiv B_{q_1}\\
B_{a=b}&=& \frac{\beta^2 p (p-1)}{64} d^{p-2} \equiv B_d
\label{AeB_1RSB}
\end{eqnarray}
where $\epsilon_{ab}$ is the matrix introduced in section
\ref{sect:rep_f_en}. For later convenience, is usuful to define:
\begin{eqnarray}
D_{\pm,\pm}&\equiv&\frac{x^2 }{2} (\eta_0\pm\theta ) (\eta_0\pm\theta )\\ \nonumber
 E_{\pm\pm}&\equiv&\frac{w x}{2}  (\eta_0 \pm \theta ) (\eta_0 \pm\theta )+\frac{x (m w+x) (\eta_1-\eta_0)(\eta_0+\frac{\pm 1\pm 1}{2} \theta + \frac{\eta_1-\eta_0}{4})}{2 m} \\ \nonumber
 G_{\pm\pm}&\equiv& \frac{x^2 (\eta_1- \eta_0)^2}{4 m^2}+\frac{w x (\eta_1-\eta_0) (\frac{\pm 1\pm 1}{2} \theta + \eta_0+\frac{3}{4}(\eta_1-\eta_0))}{m}\\ \nonumber
  &+&\frac{w^2}{2}\left[(\eta_0\pm  \theta)(\eta_0\pm \theta)+((\pm 1\pm 1)\theta +2 \eta_0)(\eta_1-\eta_0)+(\eta_1-\eta_0)^2\right]\\ \nonumber
\end{eqnarray}
and some linear combinations:
\begin{eqnarray}
 E_0^{\pm\pm} &\equiv& D_{\pm\pm} + m E_{\pm\pm}= 
 \frac{x(x+mw)}{8}(\eta_1 + \eta_0 \pm 2\theta)(\eta_1 +\eta_0 \pm 2\theta) \nonumber \\
 G_0^{\pm\pm} &\equiv& D_{\pm\pm} + 2 m E_{\pm\pm} + m^2 G_{\pm\pm} = \frac{(x+mw)^2}{2}(\eta_1 \pm \theta)(\eta_1 \pm \theta) \nonumber \\
 G_1^{\pm\pm} &:=& 2 E_{\pm\pm} + m G_{\pm\pm} \nonumber 
\end{eqnarray}

\subsection*{Projection on subspace $S_1$}

\subsubsection*{$S_{1,0}$\\}
In the subspace $S_{1,0}$, defined by $$\delta r_{a a}=0 , \quad
(\epsilon ~\delta q)_{ab}=(\delta q~ \epsilon)_{ab}=0, \quad \delta
q_{ab} \epsilon_{ab}=0$$ with $dim(S_{1,0})= 3
\frac{n}{2m^2}(n-m)(m^2-2m)$, the Hessian reduces to:
\begin{equation}
\begin{array}{c}
H_{1,0} \equiv 
\left(
\begin{array}{l l l}
D_{++}  & D_{++}- \frac{x}{2}  &   -2 D_{+-}   \\
 D_{++}- \frac{x}{2} & D_{++}  & -2 D_{+-}  \\
 -2 D_{+-} & -2 D_{+-}  &  4 D_{--} +x  \\
\end{array} \right)
\end{array}
\end{equation}
Eigenvalues of $H_{1,0}$
 \begin{eqnarray}
\hspace*{-2cm}
\Lambda_{1,0}^{(1)}&=&\frac{x}{2}
\\
\hspace*{-2cm}
\Lambda_{1,0}^{(2,3)}&=& 2 D_{--}+ D_{++}+ \frac{x}{4}
 \pm \sqrt{\left(2 D_{--}- D_{++}+ \frac{3x}{4}\right)^2 +8 D_{+-}^2  }
\end{eqnarray}
are always definite positive.

\subsubsection*{$S_{1,1}$}
In the subspace $S_{1,1}$, defined by: $$\delta r_{a a}=0 , \quad
(\epsilon ~\delta q)_{ab}=(\delta q~ \epsilon)_{ab}=0, \quad \delta
q_{ab} (1-\epsilon_{ab})=0$$ with $dim(S_{1,1})= 3\frac{n}{2}(m-3)$,
\begin{equation}
\begin{array}{c}
H_{1,1} \equiv 
\left(
\begin{array}{l l l}
D_{++} - B_{q_1} & D_{++}-\frac{x}{2} - B_{q_1}  & - 2 D_{+-} -2B_{q_1}  \\
 D_{++}- \frac{x}{2} -B_{q_1} & D_{++} - B_{q_1}  & -2 D_{+-} -2B_{q_1}  \\
 -2 D_{+-} - 2B_{q_1} & -2 D_{+-} -2 B_{q_1} &  4 D_{--} +x -4B_{q_1} \\
\end{array} \right)
\end{array}
\label{H_11}
\end{equation}
the eigenvalues are: 
\begin{eqnarray}
\hspace*{-2cm}\Lambda_{1,1}^{(1)}&=&\frac{x}{2}\\
\hspace*{-2cm}\Lambda_{1,1}^{(2,3)}&=& 2 D_{--}+ D_{++}+ \frac{x}{4} - 3B_{q_1}
\\
\nonumber
&&\qquad \pm \sqrt{\left(2 D_{--}- D_{++}+ \frac{3x}{4}-B_{q_1}\right)^2 +8 (D_{+-}-B_{q_1})^2  }
\end{eqnarray}
Since $\Lambda_{1,0}^{(1,2,3)}$ and $\Lambda_{1,1}^{(1)}$ are positive
definite, the stability condition is expressed by
$\Lambda_{1,1}^{(3)}> 0$, which reduces to: 
\begin{equation}
\eta_1>(p-1) \eta_0
\end{equation}
which concides with the dynamical condition, Eq. (\ref{MAX_COMPL}).

\subsection*{Projection on subspace $S_2$}
\subsubsection*{$S_{2,0}$\\}
In the subspace $S_{2,0}$, orthogonal to $S_{1,0} \cup S_{1,1}$ defined
by: $$\delta r_{a a}=0 , \quad (\epsilon ~\delta q
~\epsilon)_{ab}=0, \quad (\delta q~\epsilon)_{ab}\epsilon_{ab}=0$$
with $dim(S_{2,0})= 3 \frac{n}{m}\left(\frac{n}{m}-1 \right)(m-1)$,
the Hessian reduces to:
\begin{equation}
\begin{array}{c}
H_{2,0} \equiv 
\left(
\begin{array}{l l l}
E_0^{++}  & E_0^{++}- \frac{2 x + m w}{4}  &   -2 E_0^{+-}   \\
 E_0^{++}- \frac{2 x + m w}{4} & E_0^{++}  & -2 E_0^{+-}  \\
 -2 E_0^{+-} & -2 E_0^{+-}  &  4 E_0^{--} +\frac{2x+m w}{2}  \\
\end{array} \right)
\end{array}
\end{equation}
The characteristic polynomial is:
\begin{equation}
 P[H_{2,0}]=\left(\Lambda-\frac{2x +m w}{4}\right)\left(\Lambda^2 - b_{20} \Lambda + c_{20}\right)
\end{equation}
with:
\begin{eqnarray}
 b_{20} &=&  \frac{3 \left[4 \theta ^2+(1+y)^2 \eta_1^2\right]}{64 y \theta^2 \eta_1^2}  \nonumber \\
 c_{20} &=& \frac{(1+y)^2}{128 y^2 \theta ^2 \eta_1^2}  \nonumber
\end{eqnarray}
with $y=\eta_0/\eta_1$, cf. Eq (\ref{f:ydef}). 
The positiveness of both $b_{20}$ and $c_{20}$ implies positive zeros for the
polynomial.

\subsubsection*{$S_{2,1}$\\}
In the subspace $S_{2,1}$, orthogonal to $S_{1,0} \cup
S_{1,1}$ and restricted by:
\begin{equation}
\nn
(\epsilon \delta r \epsilon)_{ab} =
\sum_{c} (\delta r_{cc} \epsilon_{ac})=0 , \quad (\epsilon \delta q
\epsilon)_{ab}=0, \quad (\delta q\epsilon)_{ab}(1-\epsilon_{ab})=0
\end{equation}
with $dim(S_{2,1})= 3 \frac{n}{m}(m-1) + n- \frac{n}{m} $. Defining the
matrix:
\begin{equation}
H_{2,1}':=
\left(
\begin{array}{c | c}
 H_{1,1} & 0 \\
 \hline
 0 &   4 D_{--} +x -4B_{q_1} \\
\end{array}
\right)
\end{equation}
the Hessian projection in $S_{2,1}$ reduces to:
\begin{eqnarray}
\label{Hess:S21}
\hspace*{-2cm}\begin{array}{c}
 H_{2,1}= H_{2,1}' 
+(m-2) \left(
\begin{array}{c c c c }
E_{++} & E_{++}-\frac{y}{4} & -2E^{+-} & -2 E^{+-} \\
E_{++}-\frac{y}{4} & E_{++} & -2E^{+-} & -2 E^{+-} \\
-2 E^{+-} & -2 E^{+-} & 4E^{--}+\frac{y}{2} & 4 E^{--}+\frac{y}{2} \\
-4 \frac{E^{+-}}{m-2} & -4 \frac{E^{+-}}{m-2} & \frac{8E^{--} +y}{m-2} & \frac{8E^{--} +y}{m-2} \\
\end{array} \right)
\end{array}
\nn
\\
\label{H_21}
\end{eqnarray}
This matrix has to be studied numerically on the static (and dynamic) 1RSB solution. 
In the whole $D,T$ plane it displays two real postive eigenvalues. The other two can be both real (and positive) or complex conjugated. 
In the region of $D,T$ values where complex eigenvalues occur they have a positive real part, apart from a small subregion next to the static (resp. dynamic) RFOT  transition line.

\subsection*{Projection on subspace $S_3$}
\subsubsection*{$S_{3,0}$\\}
In the subspace $S_{3,0}$, orthogonal to all other subspaces and defined by:
\begin{equation}
  \delta r_{aa}=0 , \quad (\epsilon ~\delta q ~\epsilon )_{ab} \epsilon_{ab} = 0
\end{equation}
 with $dim(S_{3,0})= 3 \frac{n}{2m} \left( \frac{n}{m}-1 \right)$, the
 Hessian reduces to:
\begin{eqnarray}
\label{Hess:S30}
\hspace*{-2cm}
\begin{array}{c}
H_{3,0} \equiv 
\left(
\begin{array}{c c c}
G_0^{++}  & G_0^{++}- \frac{ x + m w}{2}  &   -2 G_0^{+-}   \\
 G_0^{++}- \frac{x + m w}{2} & G_0^{++}  & -2 G_0^{+-}  \\
 -2 G_0^{+-} & -2 G_0^{+-}  &  4 G_0^{--} + 2x+m w  \\
\end{array} \right)
\end{array}
\nn
\\
\end{eqnarray}
The characteristic polynomial is:
\begin{equation}
 P[H_{3,0}]=\left(\Lambda-\frac{2x +m w}{4}\right)(\Lambda^2 - b_{30} \Lambda + c_{30})
\end{equation}
with:
\begin{eqnarray}
 b_{30} &=& \frac{3}{16} \left(\frac{1}{\theta ^2}+\frac{1}{\eta _ 1^2}\right)  \nonumber \\
 c_{30} &=&  \frac{1}{32 \theta^2 \eta_1^2} \nonumber
\end{eqnarray}
the positiveness of both $b_{30}$ and $c_{30}$ implies positive zeros for the polynomial.

\subsubsection*{$S_{3,1}$\\}
In the subspace $S_{3,1}$, orthogonal to all other subspaces and
resricted by:
\begin{equation}
\nn
(\epsilon~ \delta q ~\epsilon )_{ab} ( 1 - \epsilon_{ab}
) = 0 
\end{equation}
 with $dim(S_{3,1})= 4 \frac{n}{m} $ the Hessian reduces to:
\begin{eqnarray}
\label{Hess:S31}
\hspace*{-2cm}
\begin{array}{c}
 H_{3,1}= H_{2,1}' + 
(m-1)\left(
\begin{array}{c c c c}
G_1^{++} & G_1^{++}-\frac{y}{2} & -2G_1^{+-} & -2 G_1^{+-} \\
G_1^{++}-\frac{y}{2} & G_1^{++} & -2G_1^{+-} & -2 G_1^{+-} \\
-2 G_1^{+-} & -2 G_1^{+-} & 4G_1^{--}+y & 4 G_1^{--}+y \\
- 4\frac{G_1^{+-}}{m-1} & -4 \frac{G_1^{+-}}{m-1} & \frac{8G_1^{--}+y}{m-1} & \frac{8G_1^{--}+y}{m-1} \\
\end{array} \right)
\end{array}
\nn
\\
\end{eqnarray}
Also this matrix has to be studied numerically, as Eq. (\ref{H_21}) and  we evaluated its values 
point by point
in the phase diagram. This matrix, as well, has two real positive definite
eigenvalues together with two complex conjugated eigenvalues (in a region of the $D,T$ plane).
 Some points, next to the RFOT transition have complex eigenvalues with negative
real part. Comparing with the study of compressibility, cf. Sec. \ref{ss:compress},
all point with a negative compressibility also have at
least two complex eigenvalues with a negative real part.  The opposite, as we
have shown in. Fig \ref{fig:PhDi_eigenvalues}, is not always true.

\end{document}